\documentclass[aps,prb,twocolumn,superscriptaddress,noshowpacs,nofootinbib]{revtex4-2}

\usepackage{amsmath}
\usepackage[unicode]{hyperref}
\usepackage{color,graphicx}
\usepackage[dvipsnames]{xcolor}
\usepackage[thinspace,thinqspace]{SIunits}
\usepackage{textcomp}

\newcommand{\LBCOx}{La$_{2-x}$Ba$_{x}$CuO$_{4}$}
\newcommand{\LBCOc}{La$_{1.875}$Ba$_{0.125}$CuO$_{4}$}
\newcommand{\tso}{\ensuremath{T_\text{SO}}}
\newcommand{\tco}{\ensuremath{T_\text{CO}}}
\newcommand{\tcoso}{\ensuremath{T_\text{CO,SO}}}
\newcommand{\tltt}{\ensuremath{T_\text{LTT}}}

\newcommand{\algl}{\ensuremath{{A_{\text{1g,1}}}}}
\newcommand{\algd}{\ensuremath{{A_{\text{1g,2}}}}}
\newcommand{\blg}{\ensuremath{{B_{\text{1g}}}}}
\newcommand{\bdg}{\ensuremath{{B_{\text{2g}}}}}

\begin{document}
	\title{Uniaxial stress study of spin  and charge stripes in \LBCOc\ by $^{139}$La NMR and $^{63}$Cu NQR} 
	
\author{I.~Jakovac}
\affiliation{Department of Physics, Faculty of Science, University of Zagreb, Bijeni\v {c}ka 32, Zagreb HR 10000, Croatia}

\author{A.~P.~Dioguardi}
\altaffiliation{Present address: Los Alamos National Laboratory, Los Alamos, New Mexico 87545, USA.}
\affiliation{Leibniz Institute for Solid State and Materials Research, Helmholtzstr. 20, D-01069 Dresden, Germany}

\author{M.~S.~Grbi\'{c}}
\email{corresponding author: mgrbic@phy.hr}
\affiliation{Department of Physics, Faculty of Science, University of Zagreb, Bijeni\v {c}ka 32, Zagreb HR 10000, Croatia}

\author{G.~D.~Gu}
\affiliation{Condensed Matter Physics and Materials Science Department, Brookhaven National Laboratory, Upton, New York 11973, USA}

\author{J.~M.~Tranquada}
\affiliation{Condensed Matter Physics and Materials Science Department, Brookhaven National Laboratory, Upton, New York 11973, USA}

\author{C.~W.~Hicks}
\affiliation{School of Physics and Astronomy, University of Birmingham, Birmingham, B15 2TT, UK}
\affiliation{Max Planck Institute for Chemical Physics of Solids, 01187 Dresden, Germany}

\author{M.~Po\v zek}
\affiliation{Department of Physics, Faculty of Science, University of Zagreb, Bijeni\v {c}ka 32, Zagreb HR 10000, Croatia}

\author{H.-J.~Grafe}
\affiliation{Leibniz Institute for Solid State and Materials Research, Helmholtzstr. 20, D-01069 Dresden, Germany}

	\date{\today} 
	\begin{abstract}
		We study the response of spin and charge order in single crystals of \LBCOc\ to uniaxial stress, through $^{139}$La nuclear magnetic resonance (NMR) and $^{63}$Cu nuclear quadrupole resonance (NQR), respectively. In unstressed \LBCOc, the low-temperature tetragonal structure onsets below $\tltt= 57$~K, while the charge order and the spin order transition temperatures are  $\tco = 54$~ K and $\tso = 37$~K, respectively.  We find that uniaxial stress along the [110] lattice direction strongly suppresses \tco\ and \tso, but has little effect on \tltt. In other words, under stress along [110] a large splitting ($\approx 21$~K) opens between \tco\ and \tltt, showing that these transitions are not tightly linked. On the other hand, stress along [100] causes a slight suppression of \tltt\ but has essentially no effect on  \tco\ and \tso. Magnetic field $H$ along [110] stabilizes the spin order: the suppression of \tso\ under stress along [110] is slower under $H \parallel [110]$ than $H \parallel [001]$. We develop a Landau free energy model and interpret our findings as an interplay of symmetry breaking terms driven by the orientation of spins.
	\end{abstract}

	\maketitle

	\section{Introduction}
    One of the leading open questions in the research of high-temperature superconductors is the relation between competing electronic orders. Even though stripe charge order (CO) is ubiquitous in cuprates, the relationship between static charge and spin order (SO) remains incompletely understood. This is partly due to the limited number of systems in which both can be studied. The other reason is that the structural, electronic, and magnetic degrees of freedom are intertwined in these orders. In \LBCOx\ (LBCO) close to $x=1/8$ doping, CO becomes pinned as the symmetry of the lattice changes from low-temperature orthogonal (LTO) to low-temperature tetragonal (LTT) at $\tltt = 57$~K. At this doping, the CO onsets below $\tco \approx 54$~K, and SO transition temperature $\tso$ reaches its maximum value \cite{Kumagai1994, Goto1994, HuckerPRB, GuguchiaPRB} of $\approx$40~K. In contrast, the bulk superconducting transition temperature ($T_c$) is strongly suppressed. $T_c$ rapidly increases for doping away from 1/8, even though the structural transition and CO/SO persist. It was initially hypothesized that the structural symmetry of the LTT phase is necessary for CO/SO to condense. However, H\"ucker et al.~\cite{HuckerPRL} have shown in \LBCOc\ under hydrostatic pressure	that CO/SO appear even when long-range LTT structural order was suppressed, which softened the structural symmetry restriction.
	Follow-up studies found that  CO was persisting in the presence of \textit{local} LTT lattice deformations
	\begin{figure}
		\includegraphics[width=0.48\textwidth]{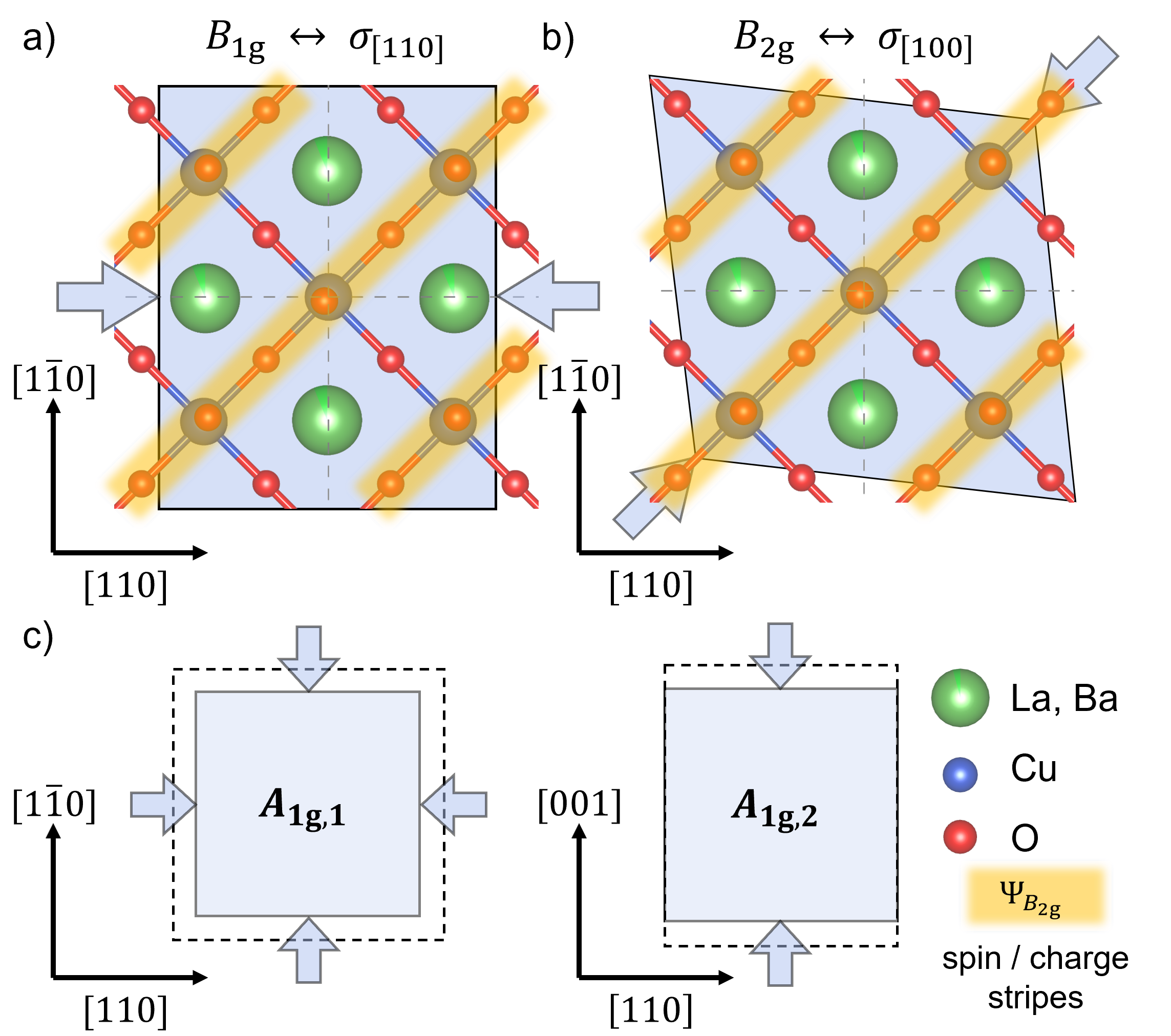}
		\caption{Schematic of characteristic  in-plane symmetry-breaking strains a) \blg\ (orthorhombic) and b) \bdg\ (rhombic), and c) symmetric, \algl\ and \algd. The unstrained lattice in the foreground illustrates how the strain is applied with respect to the CO and SO parameter $\Psi_\bdg$, structural symmetry-breaking order parameter $\Phi_\bdg$ (octahedral tilts). \blg\ and \bdg\ denote the irreducible representations of $D_\text{4h}$ point group. Strain directions are expressed in the principal axes of the HTT phase (see text).} \label{fig:symmetry}
	\end{figure}
	\cite{FabbrisPRB, FabbrisHPR}, which put the role of structural symmetry in this compound in focus again. Similarly ambiguous connection of stripe order to the structure is also seen in other rare-earth doped systems (La$_{2-x-y}$RE$_y$Sr$_x$CuO$_4$, RE = Eu, Nd), where CO is known to appear within the LTT phase or at least close to \tltt \cite{CrawfordPRB, SimovicEPL, FabbrisPRB, FabbrisHPR, HuckerPRL, GuguchiaPRB, BoylePRR, ArumugamPRL, TakeshitaJPSJ}. A complex interplay of disorder, symmetry, and electron correlations completely changes how CO/SO appears.\\
	\indent In this paper, we use nuclear magnetic and quadrupolar resonance  (NMR/NQR) to systematically study the phase diagrams of SO, CO, and LTT structure onset in archetypal stripe compound \LBCOc, controlled by in-plane uniaxial stress ($\sigma$) in [100] and [110] directions (see Fig.~\ref{fig:symmetry}). 
	It has previously been reported with $\mu$SR that stress approximately along [110] rapidly suppresses \tso\ in La$_{1.895}$Ba$_{0.115}$CuO$_4$. Here, we find \tso\ is also strongly suppressed by stress along [110] in \LBCOc\ although the SO is more robust than for $x=0.115$: larger stress is required. As magnetic field ($H$) is required to carry out the NMR measurements, we find that \tso\ is suppressed more slowly for $H \parallel [110]$ than $H \parallel [001]$, pointing to a non-trivial interplay of spin direction and lattice symmetry.
		The CO shows the equivalent response to external stress, and the onset temperature \tco\ is strongly suppressed by $\sigma_{[110]}$. However, \tltt\ shows only a mild response to applied stress. As a result, $\sigma_{[110]}$ causes \tco\ to separate from \tltt, with $\tltt-\tco \approx 21$~K at maximal induced strain. This unexpected result resolves how structural symmetry affects the formation of stripe order in cuprates. We discuss these findings as an interplay of symmetry-defined terms in a self-developed Landau free energy model that simultaneously shows a good agreement with earlier data dependence of $T_\text{CO,SO}$ to hydrostatic pressure.\\
	\indent The paper is organized as follows: in Sec.~\ref{EM}, we explain the experimental methods used in the study; in Sec.~\ref{R}, we present the results of La NMR and Cu NQR; in Sec.~\ref{LFE}, we present the Landau Free Energy model developed to analyze our results; in Sec.~\ref{D}, we discuss our findings and summarize in Sec.~\ref{Con}.

	\section{Experimental methods}
	\label{EM}
	The \LBCOc\ single crystal samples were grown with the traveling solvent floating-zone method described in Ref.~\cite{GUJCG}. Samples were first properly aligned by Laue scattering and cut along the specific crystallographic directions in the high-temperature tetragonal (HTT) phase. The typical sample size used in the experiment was $4\times 1 \times 0.5$~mm$^3$, where the longest dimension was either [100] or [110], and the shortest was along [001]. By [110], we denote the direction along the diagonal of the CuO$_2$ square lattice with Cu in the corners, and by [100] the direction along the Cu--O bond (see Figs.~\ref{fig:symmetry} ~a) and b)). When a symmetry-breaking stress is exerted on the sample, neither phase remains strictly orthorhombic (above $\tltt$) nor tetragonal (below $\tltt$). However, we will continue using the same notation to prevent potential confusion and to stay consistent with the notation used in other articles on the topic. We characterized both samples by SQUID magnetometry in low magnetic fields of 20~Oe. They showed the same behavior below 40~K and a bulk $T_c$ of about 5.5~K as in Ref.~\cite{TranquadaPRB08}.\\
	\indent NMR data on lanthanum (spin $I=7/2$, $\gamma_n /2 \pi= 6.0146$ MHz/T) were collected on a central (+1/2 $\leftrightarrow$ --1/2) transition of the $^{139}$La spectra using a Tecmag spectrometer with a Hahn echo pulse sequence $\pi/2-\tau-\pi$. Typical $\pi/2$ pulse length was 0.5~$\mu$s and $\tau= 17$~$\mu$s, while pulse power was $0.5$~W. With the magnetic field of 7~T the spin-lattice relaxation rates $T_1 ^{-1}$ were measured at frequency $\omega_L = 42.18$~MHz. $T_1 ^{-1}$ relaxation rates were determined by a saturation-pulse recovery sequence, after which the data was fit to a relaxation curve \cite{Narath1967,MacLaughlinPRB} for $I=7/2$: $f(t) = (1/84)e^{-(t/T_1)^{s}}  + (3/44) e^{-(6t/T_1)^{s}} + (75/364)e^{-(15t/T_1)^{s}} + (1225/1716)e^{-(28t/T_1)^{s}}$. The phenomenological stretching exponent $s$ gives insight into the distribution of the relaxation times $T_1$. The $s \geq 0.5$ implies the Gaussian $T_1$ distribution on a logarithmic scale with FWHM across an order of magnitude and $T_1 \approx T_{1,\text{median}}$. When $s < 0.5$, the distribution widens drastically, and the fitted $T_1$ no longer represents the distribution median \cite{Johnston2006}. NQR data on $^{63}$Cu were collected on the high-frequency signal (the so-called B-line) from Cu sites near the dopant Ba ions~\cite{SingerPRL}. Since in \LBCOc\ the B-line is well separated from the low-frequency A-line, it can be analyzed directly without additional spectral deconvolution. To acquire the signal, we employed the method reported in a previous work~\cite{PelcPRB}, using a Hahn echo with a typical $\pi/2$ pulse length of 0.7~$\mu$s and $\tau= 4$~$\mu$s. Since the Cu NQR line intensity rapidly diminishes~\cite{HuntPRB,GrafePRL, SingerPRB1999,CurroPRL2000,PelcNC} at the onset of CO, this was utilized to determine \tco\ .\\
	\indent To induce strain, we employed a uniaxial strain cell described in~\cite{Hicks2014} (partly shown in Fig.~\ref{fig:sample}). The applied stress was varied by applying voltages $V = \pm 200$~V, with which we were able to induce a strain up to $\varepsilon \approx 1\%$, depending on the sample orientation and dimensions. To deduce the applied stress ($\sigma$~[GPa]) from the measured strain ($\varepsilon$~[$\%$]), we used the cuprate elasticity data from \cite{NoharaPRB}. Technical details are shown in Appendix~A.
	\begin{figure}[h!]  
	\includegraphics[width=0.433\textwidth]{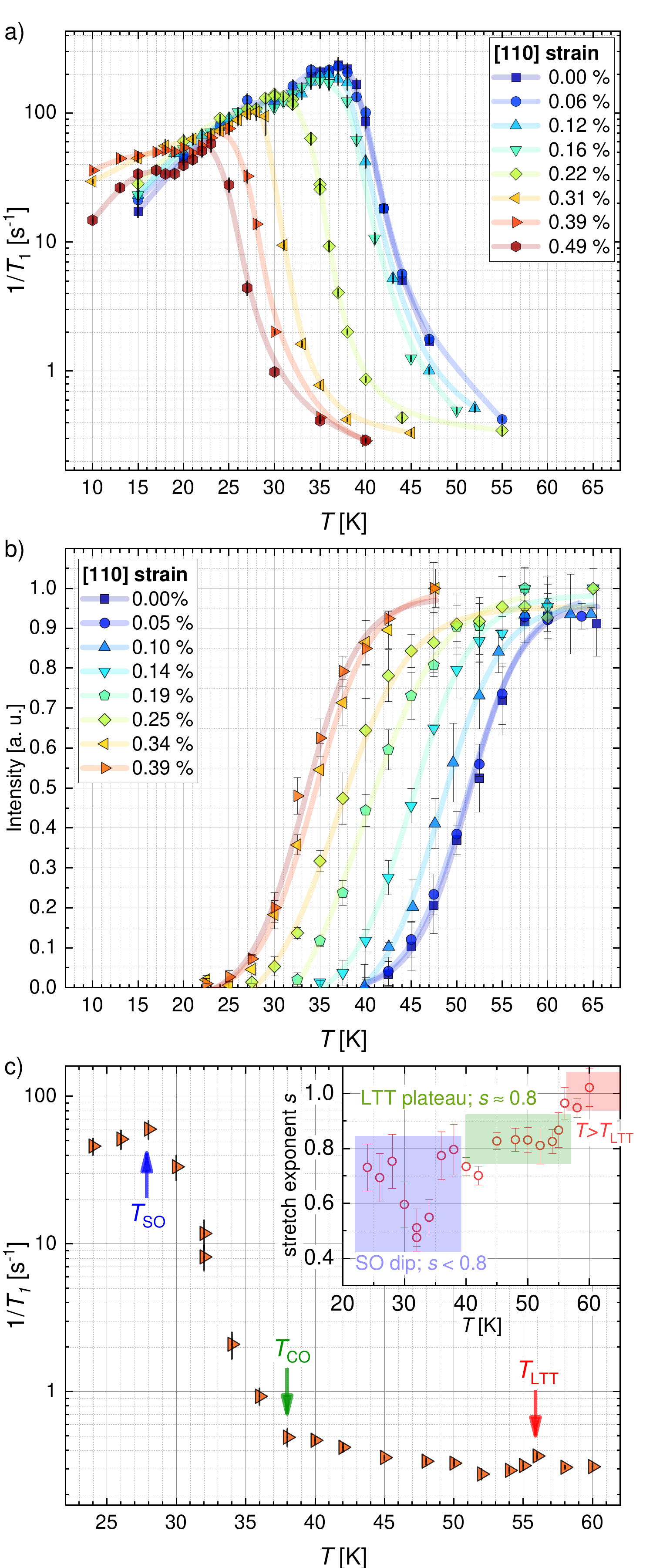}
	\caption{a) Temperature dependence of $^{139}$La spin-lattice relaxation rate $^{139}T_1 ^{-1}$ for $H\parallel  [001]$  and stress applied in [110]. Maximum in $T_1 ^{-1}$ coincides with \tso.  Lines are guides to the eyes. b) Normalized intensity of $^{63}$Cu B-line measured by NQR under $\varepsilon_\text{[110]}$ strain. The lines are fitted to a phenomenological function $I(T)$ (see text). c)  $T_1 ^{-1}$ measurement at uniaxial strain $\varepsilon_\text{[110]} = 0.4\%$ in a wide temperature range shows an anomaly at $T_\text{LTT} \approx 56$~K. In the inset, stretch exponent $s$ drops slightly at \tltt, then to $s = 0.5$ close to $T_\text{SO}$. The legend shows the values of measured $\varepsilon_\text{[110]}$ and $\varepsilon_\text{[100]}$ strain.}
	\label{fig:NMR&NQR}
	\end{figure} 

	\section{Results}
	\label{R}
\begin{figure}[t]
		\includegraphics[width=0.5\textwidth]{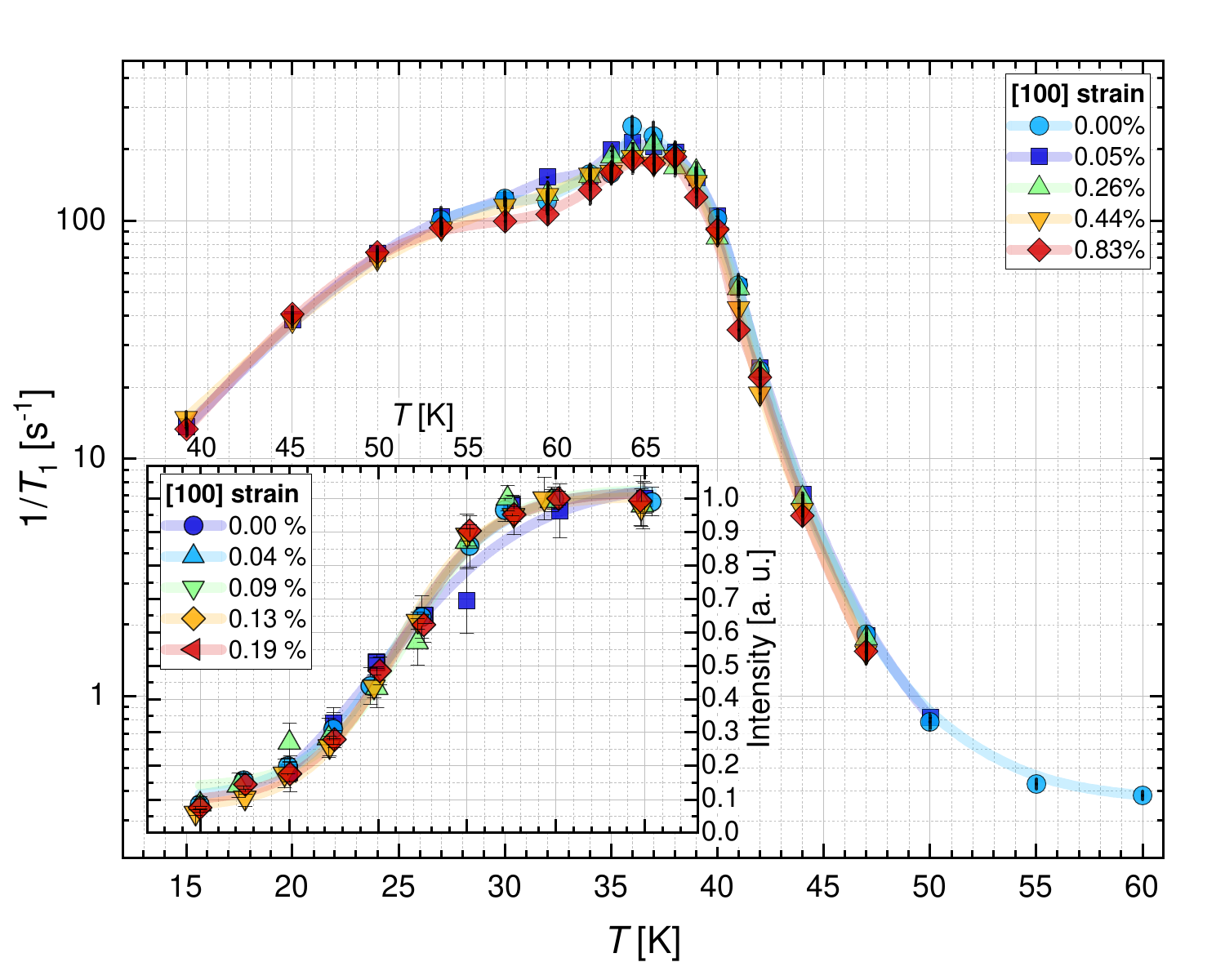}
		\caption{Temperature dependence of $^{139}T_1 ^{-1}$ measured with stress applied along [100] and $H\parallel[001]$. Lines are guides to the eyes. The inset shows the NQR measurements of copper B-line at $\sigma_{[100]}$. Lines are fits to the phenomenological function $I(T)$ (see text). The legend shows the values of induced strain. }
		\label{fig:T1_100}
	\end{figure}

\begin{figure*}[t]
		\includegraphics[width=0.99\textwidth]{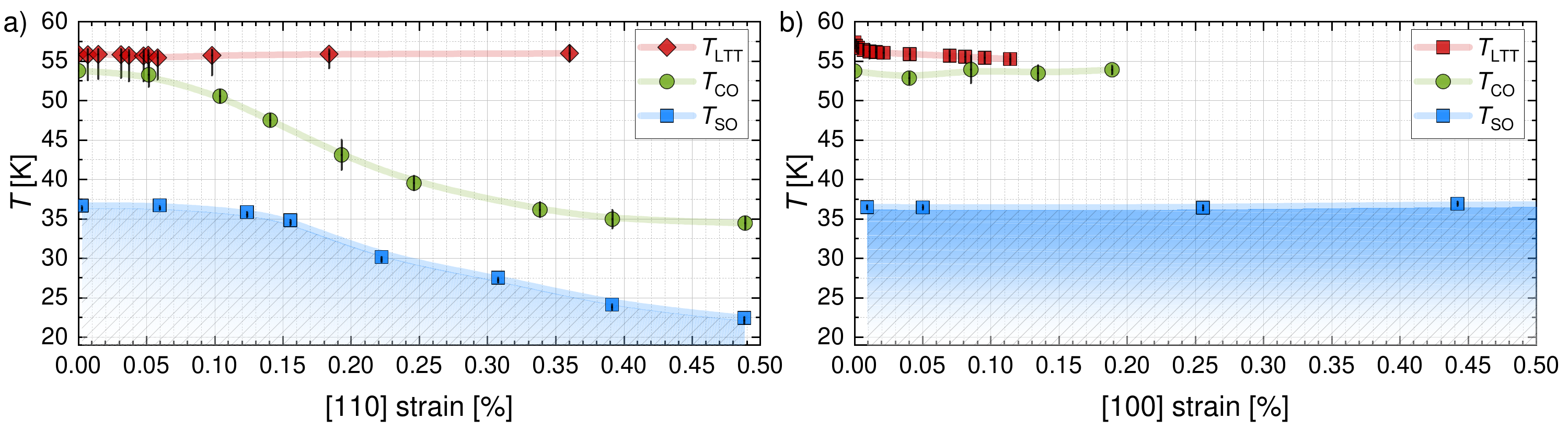}
		\caption{Strain-temperature phase diagram for $H\parallel c$ and stress applied a) along the [110] direction, and b) along the [100] direction. The data points are extracted from 
		the data of Figs.~\ref{fig:NMR&NQR} and~\ref{fig:T1_100} and show that $\sigma_{[110]}$ reduces $\tso$ (blue) and \tco\ (green),  even though the onset of LTT structural phase remains the same (red). However, $\sigma_{[100]}$ does not change $\tso$ at all, while $\tltt$ shows a mild drop. Lines are guides to the eyes. }
		\label{fig:phase_diagram}
	\end{figure*}

	\indent In Figs.~\ref{fig:NMR&NQR}~a) and c) we present the measured temperature dependence of the spin-lattice relaxation rate $T_1 ^{-1}$ for stress applied along the [110].
	$\varepsilon_{110}$ denotes the induced strain along the $[110]$ direction, obtained under $\sigma_{110}$. Poisson's-ratio expansion in the transverse directions is implied.
	 The magnetic field of 7~T was oriented along the crystal [001] axis. In the unstrained sample  $T_1 ^{-1}$ starts to increase below 55~K, as CO onsets. With cooling, critical slowing down of spin fluctuations cause $T_1^{-1}$ to increase by three orders of magnitude before reaching a maximum value at \tso ~=~37~K 
	 at zero strain. 
	 With further cooling, $T_1 ^{-1}$ slowly decreases as the fluctuations of the SO 
	 continue to slow down. For $H\parallel c$, such temperature dependence of $^{139}$La $T_1 ^{-1}$  has been shown~\cite{SuhPRBR,CurroPRL2000,BaekPRB} to deviate from the  Bloembergen-Pound-Purcell (BPP) mechanism~\cite{BPP}: $T_1 ^{-1} (T) = \gamma^2 h_0 ^2 \tau_c (T)/(1+\omega_L ^2 \tau_c ^2 (T))$, where $h_0$ is the local field fluctuating at the nuclear site, $\gamma$ is the gyromagnetic ratio, $\omega_L=\gamma H$ is the nuclear Larmor frequency, and $\tau_c (T) = \tau_{\infty} \exp(E_a/k_B T)$ is the electron relaxation time $\tau_ c$ with an activation energy $E_a$.  $T_1 ^{-1} (T) $ is somewhat better described by the extended BPP model where $E_a$ is introduced with a normal distribution of values of a typical width 80~K, although it still cannot account for the complete behavior. The distribution of $E_a$ is typically explained by the intrinsic level of disorder in the cuprates. Nevertheless, we shall discuss some aspects of the observed $T_1 ^{-1} (T) $ dependence (however, only qualitatively) through BPP model parameters since despite its limitations no better model is currently available.\\
	\indent When stress is applied along [110], for measured $\varepsilon_\text{[110]}$ strain values larger than $0.1~\%$ ($\approx$180~MPa), \tso\ shifts to lower temperatures. Also, the peak value of $T_1^{-1}$ at \tso\ decreases.
	 The width of the SO transition does not broaden, even at the highest stress value where $\tso$ is reduced by more than 35$\%$, indicating a high level of strain homogeneity, and no increase of the $E_a$ values distribution as the sample is compressed. For temperatures below $\tso$, we see that the relaxation values under stress are not simply shifted like those for $T > \tso$, but that the values smoothly connect to the $T_1 ^{-1} (T)$ dependence measured at zero stress, so that $T_1 ^{-1} (T,\varepsilon_\text{[110]})$ remain practically unchanged down to 20~K. Within the BPP model, this would indicate that the electronic fluctuation time $\tau_c$ 
	 is unaffected (or reduces together with $h_0$) by stress, and is determined by the absolute temperature value $T$ rather than $T - \tso$. This is not what is typically observed with the suppression of a magnetic transition by doping or strain. One would expect that (e.g.  Fig.~3 in~\cite{DiguardiPRB}, or Fig.~6 in~\cite{NingPRB}) as $\sigma_\text{[110]}$ destabilizes SO, an increase in spin fluctuations would increase $T_1 ^{-1} $ for $T \le \tso$. Current behavior indicates a complex relationship between $H$ and fluctuations of stripe SO.\\
	 \indent To characterize the response of CO to $\varepsilon_{[110]}$, we measured the temperature dependence of the integrated intensity of the high-frequency copper NQR signal (B-line) in the vicinity of \tco, shown in Fig.~\ref{fig:NMR&NQR} b). It has been well established that the intensity of the B-line $I(T)$ reduces with the onset of CO due to the effect known as wipeout~\cite{HuntPRB,GrafePRL, SingerPRB1999,CurroPRL2000,PelcNC}. Recently, it was shown that the wipeout in \LBCOc\ is caused by incoherent spin fluctuations and the increase of NQR linewidth~\cite{PelcPRB}. 
	 To analyze our data systematically, every measured $I(T)$ dependence was corrected for temperature and then fitted to a simple phenomenological function $(\tanh((T-T_{h})/w)+1)/2$, where $T_h$ is the mid-transition temperature and $w$ is the width of the transition. Clearly, $T_h$ is related to CO onset temperature as $\tco = C T_h$, where the constant $C=1.055$ is set by the \tco\ value at zero strain ($\tco =54$~K). The value of $C$ was kept the same for all strains as the width $w$ and the shape of the transition do not change with strain. Fig.~\ref{fig:NMR&NQR} b) shows that the $\tco$ is strongly suppressed by $\varepsilon_{[110]}$.\\\
	 \indent When the applied stress is sufficient to separate \tso\ and \tco\ from \tltt, one can observe all the characteristic temperatures in $T_1 ^{-1} (T)$ measurements alone. For example, in Fig.~\ref{fig:NMR&NQR} c) we show $T_1 ^{-1} (T)$ dependence in a wide temperature range at $\varepsilon_{[110]} = 0.4$~$\%$ where \tso\ and \tco\ are easily noticeable. The small peak structure close to 56~K is attributed to \tltt, which we have determined independently and will discuss it later in the text. This measurement of $T_1 ^{-1} (T)$ was done on a different sample of the same doping, and as we can see the characteristic temperatures match those determined previously, which shows a high degree of reproducibility.\\
	\indent The behavior of $T_1 ^{-1}$ under $\sigma_\text{[110]}$ is in stark contrast to the one set by stress along [100] ($\sigma_\text{[100]}$) shown in Fig.~\ref{fig:T1_100}. Here, $T_1 ^{-1} (T)$ is essentially unaffected, even at the highest stress values. Hence, \tso\ does not change with $\varepsilon_{[100]}$. From the inset of the figure we can see that this strain direction does not affect \tco, either.\\
	\indent With $T_1 ^{-1}$, we also measured the spectral features of the $^{139}$La central transition (shown in Appendix~B) which showed no anomalous change in linewidth or shape with temperature and stress in the region of our measurements. Hence, we conclude that samples have only undergone elastic deformation without reaching a plastic regime or cracking. Furthermore, the distribution of $T_1$ times, characterized by the stretching exponent $s$ of the relaxation curves, shows the characteristic behavior observed in earlier studies~\cite{BaekPRB}.\\
	\indent The suppression of \tso\ by $\epsilon_{[110]}$ is similar to the one reported by $\mu$SR on an $x=0.115$ doped sample, for stress along a specific direction aligned at an angle of $30^{\circ}$ relative to the Cu--O bond \cite{GuguchiaPRL}. There, the authors reported a drop of $\tso$ values down to 30~K for $\sigma\approx$ 40~MPa, after which it reached a saturated value that barely changes up to the highest stress value of 90~MPa. However, at 1/8 doping the SO is more robust~\cite{HuckerPRB, GuguchiaPRB,SchottThesis}, and this is why larger stress is needed to equally suppress \tso. Our results reveal that the major effect of SO suppression actually comes from  stress along [110] direction.\\
	\indent To check how stress influences the LTO-LTT transition, we combined the measurements of $T_1 ^{-1} (T)$ and the data of voltage and capacitance measured at the strain cell. By lowering the temperature across $\tltt$, a clear anomaly is seen in displacement (Fig.~\ref{fig:displacement} in Appendix~C), caused by the change in compressibility across the structural transition~\cite{NoharaPRB}. The anomaly is small enough not to influence the overall value of applied stress but remains within the resolution of our measurement setup. As mentioned earlier, the  $\tltt (\varepsilon)$ dependence is also confirmed by measurements of $T_1^{-1}$, which shows a small peak at \tltt. Similar behavior has been observed~\cite{BaekPRB} at the HTT/LTO structural transition, and at the LTO/LTT transition in La$_{1.65}$Eu$_{0.2}$Sr$_{0.15}$CuO$_4$~\cite{SimovicEPL}. We found no noticeable effect on the onset of the LTT phase with stress applied along [110], as is shown in Fig.~\ref{fig:phase_diagram}~a). However, stress along the [100] direction causes a slow but definite suppression of $\tltt$. This is qualitatively similar to what was observed~\cite{BoylePRR} in {La$_{1.475}$Nd$_{0.4}$Sr$_{0.125}$CuO$_{4}$}, albeit of smaller size, since there $\varepsilon_{[100]}$ strain of $\approx$0.046$\%$ reduced $\tltt$ from 63~K to 34~K. A reason could be that the system is close~\cite{Crawford1991, CrawfordPRB, AxeJLTP} to a triple structural transition point rendering \tltt\ more susceptible to external stress.
		\\
    \indent From these results, we generate the (in-plane)-stress controlled phase diagrams depicted in Fig.~\ref{fig:phase_diagram}. To the best of our knowledge, this is the first time that one has traced the behavior of all three temperatures under stress. For stress $\varepsilon_{[110]}$ above $ \approx 0.06 \%$, \tco\ separates from \tltt\ and $\tltt - \tco$ reaches 21~K at maximum strain -- a dramatic change in the behavior reminiscent of the situation in La$_{1.8-x}$Eu$_{0.2}$Sr$_{x}$CuO$_4$ where $\tltt \approx 130$~K and \tco\ reaches 80~K at $x = 0.125$ doping~\cite{FinkPRB}.  Up to this point it was not possible to achieve a similarly large difference  between \tltt\ and \tco\ in another system at 0.125 doping. These data show that one can indeed separate them by inducing the strain of a specific direction.\\
    	 				\begin{figure}[b]
    	\includegraphics[width=0.5\textwidth]{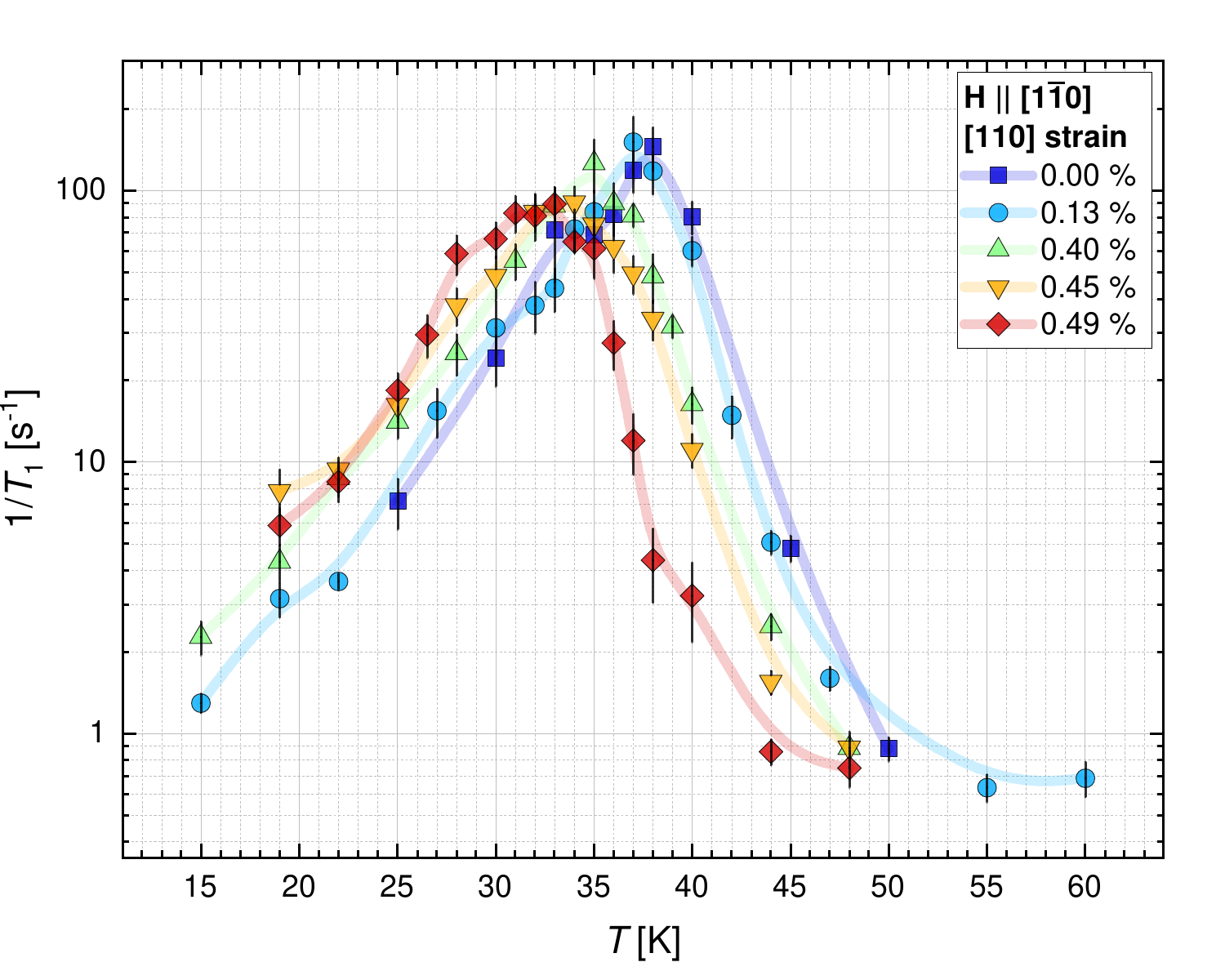}
    	\caption{Temperature dependence of $^{139}T_1 ^{-1}$ measured with stress $\sigma_{[110]}$ and $H\parallel[1\overline{1}0]$. Lines are guides to the eyes. The suppression of $\tso$ is greatly reduced. The legend shows the values of measured strain.}
    	\label{fig:T1_110_H110}
    \end{figure}
	\indent Looking back on $T_1 ^{-1}$ data in Fig.~\ref{fig:NMR&NQR}~a), in an earlier study~\cite{BaekPRB} it was found that $^{139}$La $T_1 ^{-1}$ shows a magnetic field-induced anisotropy connected to the relative orientation of spins~\cite{HuckerPRB2008} in the SO stripes with respect to the external magnetic field. In particular, in the SO state, $T_1 ^{-1}$ is approximately an order of magnitude larger for $H \parallel [001]$ in comparison to $H \parallel [110]$ (or [1$\overline{1}$0]). This difference is not caused by the anisotropic hyperfine coupling since it would then be visible even in the paramagnetic state, but rather the anisotropy reflects the property of the SO state. As was mentioned earlier, the lack of increase of $T_1 ^{-1}$ below \tso\ in Fig.~\ref{fig:NMR&NQR}~a) also shows an unusual relationship between spin fluctuations and magnetic field. To further clarify the nature of this anisotropy, we applied stress again along the [110] direction, but this time with $ H \parallel [1\overline{1}0]$. The results are shown in Fig.~\ref{fig:T1_110_H110}: for the unstressed sample, we reproduce the $T^{-1}_1$ values within the SO phase from~\cite{BaekPRB}. What is surprising, though, is that reorientation of the magnetic field drastically reduces the stress-driven suppression of $\tso$. With $H \parallel [1\overline{1}0]$, $\tso$ is reduced to only 32~K (which is $\Delta T\approx 5$~K from zero-stress value) at a $[110]$ strain of 0.49~$\%$ ($\approx$0.9~GPa). This change in $\tso$ corresponds to an overall rate of 10.2~K$/\%$ ($\approx$5.63~K/GPa), which is significantly less than 27.5~K$/\%$ ($\approx$15.2~K/GPa) obtained for $H\parallel [001]$. Clearly, the magnetic field along $[1\overline{1}0]$ reduces the effect of stress and acts as a stabilizing factor to stripe SO. 
	 This surprising result, seemingly unique to LBCO, has been implied previously \cite{HuckerPRB2008,BaekPRB}, but in this study it is directly revealed.\\
	\indent Another observation can be made from Fig.~\ref{fig:T1_110_H110} for $\varepsilon_\text{[110]}>0.13~\%$: in addition to the gradual shifting of $\tso$ to lower temperatures, it can be seen that the $T_1 ^{-1}$ values (i.e. spin fluctuations) increase for $T<\tso$, as is expected for suppressed magnetic order. Hence, spin fluctuations now seem to depend on $T-\tso$. This would indicate that the unusual anisotropy of the SO fluctuations persists even under stress.\\
		\indent We have not explored how magnetic field influences CO, since NQR measurements (performed in zero magnetic field) allow us to isolate the copper signal for a specific doping environment (B-line). When the magnetic field is applied, the NMR lines start to overlap, and it is no longer simple to assign changes in the spectra to a specific phenomenon of the stripe physics.

	\section{Free energy model}	
	\label{LFE}
	\indent To address the markedly different strain dependencies of the onset temperatures $\tltt$ and $\tcoso$, we consider a simple Landau free energy (LFE) model. A similar approach has led to the development of the linear two-component order parameter model \cite{PlakidaArticle} to explain the doping dependence of $\tltt$ in LBCO \cite{IshibashiJPSJ}, stiffness constant softening observed in ultrasound experiments \cite{MiglioriPRL}, and the out-of-plane component of magnetic moment in certain cuprate systems \cite{HuckerPRB2004, Keimer1993}. Although such a two-component approach was prevalent, it lacked the higher-order contributions necessary to model the response to symmetry-breaking in-plane strains. Thus, the strain-related research on the iron pnictides shifted the focus to the simpler, symmetry-defined, LFE models \cite{TogoPRB2008, LuPRB, ChuScience}, which helped to elucidate how the nematic order in iron pnictides couples to the symmetry-breaking strains. We can apply the same arguments to characterize the observed $\tcoso$ suppression in \LBCOc.\\
	\indent First, we focus on the SO transition revealed by the $^{139}$La $T_1 ^{-1}$  data. In the LTT phase, the  \LBCOc \ crystal point group is tetragonal $D_\text{4h}$; however, owing to the octahedral tilts along $\pm$[100] crystallographic axes, for a single CuO$_2$ layer, the in-plane symmetry is reduced. We model the LTT phase by introducing the structural order into the $D_\text{4h}$ symmetric planes. In the case of LBCO, both structural ($\Phi_{B_\text{2g}}$) and spin ($\Psi_{B_\text{2g}}$) order transform as the \bdg\ irreducible representation  of the unstrained sample's point group $D_\text{4h}$. One can readily use the same model to describe the observed suppression of $\tco$, since both spin and charge order possess the same symmetry.
	\\
	\indent The strain tensor components can be written as an in-plane symmetric ($\varepsilon_\algl = \frac{1}{2}\left(\varepsilon_{xx} + \varepsilon_{yy}\right)$, $\varepsilon_\algd = \varepsilon_{zz}$) and antisymmetric ($\varepsilon_\blg = \frac{1}{2}\left(\varepsilon_{xx} - \varepsilon_{yy}\right), \varepsilon_\bdg = \varepsilon_\text{xy}$) linear combination~\cite{IkedaPRB}. The out-of-plane shear strain components $\varepsilon_{xz}$ and $\varepsilon_{yz}$, which form a two-dimensional $E_g(1,2)$ representation of the group, are absent in our measurements and will be omitted from the model. The minimal LFE model is given by
	\begin{equation}
		F = F_\Psi + F_{\Psi\varepsilon} + F_{\Psi\Phi} + F_{\varepsilon}, 
		\label{eq:LFE}
	\end{equation}
	where $F_{\Psi} =  \Psi_{B_\text{2g}}^{2} a \left(T - T_\text{CO,SO}\right) + \Psi_{B_\text{2g}}^{4} b/2$ are the usual LFE terms ($a, b>0$) which lead to the second order phase transition, $F_{\Psi\varepsilon}$ and $F_{\Psi\Phi}$ are spin/charge-strain and spin/charge-structure coupling terms, respectively, and $F_{\varepsilon}$ is the elastic energy. To the lowest order in $\Psi_{B_\text{2g}}$ we have:
	\begin{align}
		F_{\Psi\varepsilon} &=  \alpha_{1}\varepsilon_{A_\text{1g,1}} \Psi_{B_\text{2g}}^{2}  + \alpha_{2} \varepsilon_{A_\text{1g,2}} \Psi_{B_\text{2g}}^{2} + \nonumber\\
		&+  \beta \varepsilon_{B_\text{1g}}^{2} \Psi_{B_\text{2g}}^{2}+ \gamma  \varepsilon_{B_\text{2g}} \Psi_{B_\text{2g}},
		\label{eq:varepsilonPsi}
	\end{align} 
	where the parameters $\alpha_1$ and $\alpha_2$ define the coupling strength to the symmetric, and $\beta$ and $\gamma$ to the antisymmetric strain. The symmetry considerations allow for a quadratic charge- and spin-structure coupling $F_{\Psi\Phi}^\text{(CO,SO)} = \delta \Phi_{B_\text{2g}}^2 \Psi_{B_\text{2g}}^2$, due to the difference between charge (spin) and structure order wavevectors. 
	Finally, the elastic energy is given by:
	\begin{align}
		F_{\varepsilon} &= \varepsilon_{A_\text{1g,1}}^{2} \left(C_{11} + C_{12}\right) + C_{33} \varepsilon_{A_\text{1g,2}}^{2}/2 + \varepsilon_{B_\text{1g}}^{2} \left(C_{11} - C_{12}\right) \nonumber \\
		&+  2 C_{13} \varepsilon_{A_\text{1g,1}} \varepsilon_{A_\text{1g,2}} + 2 C_{66} \varepsilon_{B_\text{2g}}^{2}.
		\label{eq:varepsilon}
	\end{align}\\
	As presented, LFE also captures the evolution of CO/SO parameter magnitude $\left| \Psi_{B_\text{2g}} \right|$ with the changes in structural order $\Phi_{B_\text{2g}}$, but this should not affect the $\tcoso$. Hence, we will only focus on \tcoso\, since we have no data to discuss the magnitude.
	
	\indent The emergence of the CO and SO induces spontaneous strains in the lattice when cooled below \tcoso\, which form a rhombic distortion, suggesting that the external rhombic [100] stress (Fig.~\ref{fig:symmetry} c)) would only lead to a finite order parameter at all temperatures \cite{IkedaPRB}, and a crossover instead of a phase transition (see Appendix~D). However, the crossover of the CO and SO transitions (detected by, e.g., temperature broadening) is not visible in our measurements in Fig.~\ref{fig:T1_100}, so we can conclude that the coupling to the rhombic strain is minimal. On the other hand, orthorhombic strain [110] breaks an additional symmetry, introducing more terms into the electronic Hamiltonian, and acts as a tuning parameter for the CO and SO transitions.\\
	\indent  Minimizing the LFE in equation (\ref{eq:LFE}) with respect to the order parameter $\Psi_{B_\text{2g}}$ exposes a functional correlation between the applied stress and both structural and spin transition temperatures. Moreover, the intricate strain-order interaction will cause the structural strain to appear in the ordered phase without external stress (see Appendix D for more details). For clarity, we have replaced all elastic constants $C_{ij}$ with the appropriate elasticity parameters (Young moduli and Poisson ratios).
	
	\section{Discussion}	
	\label{D}
\begin{figure}[t]
		\includegraphics[width=0.49\textwidth]{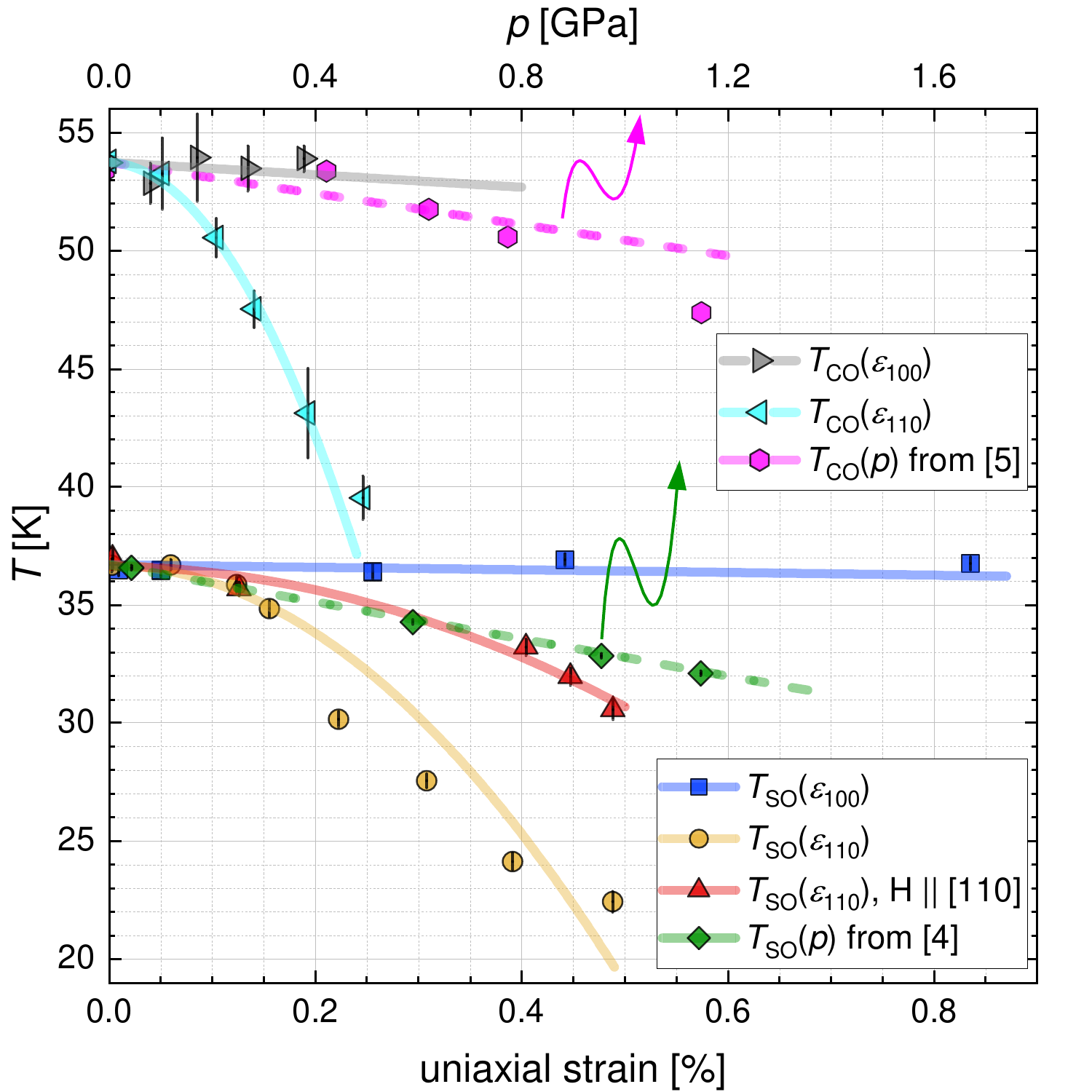}
		\caption{\tso\ and \tco\  suppression induced by different strains (lower axis) and hydrostatic pressure $p$ (top axis). The points are experimental data showing either strain dependence from this work, or hydrostatic pressure dependence from Ref.~\cite{GuguchiaPRB} ($p>1.2$~GPa data are omitted for clarity) and Ref.~\cite{HuckerPRL}. Full thick lines mark curves fitted to the LFE model (see text), while dashed lines are predictions of $T_\text{CO,SO}(p)$ \textit{calculated} from the model. Curvy arrows indicate hydrostatic pressure data (measured and calculated) should be read on the top axis.
		}
		\label{fig:model_fit}
	\end{figure}

	We take applied stress as a control parameter to uncover the \tcoso\ dependence on the measured strain. While below the \tcoso\ the elastic constants are renormalized by the emergent order \cite{plakida}, above the transition temperature, the strain on the sample depends only on its elastic properties. For stress along [100], the dependence of the \tcoso\ is then proportional to the symmetric stress contributions:
	\begin{equation}
		\frac{\partial T_\text{CO,SO}}{\partial \sigma_\text{[100]}} = \frac{\alpha_1(1-\nu_\text{in})-2\alpha_2\nu_\text{out}}{2Y_\text{[100]} a},
		\label{eq:tso100}
	\end{equation}
	where $\nu_\text{in}$ and $\nu_\text{out}$ are in-plane and out-of-plane Poisson ratios, respectively, and $Y_\text{[100]}$ is a Young modulus along the [100] axis.
	The lack of any observable change in the $\tso$ and $\tco$ measurements under $\sigma_{[100]}$ suggests that the two symmetric stress contributions in (\ref{eq:tso100}) are either small or exactly cancel each other out. In contrast, when applying [110] stress to the sample, from~(\ref{eq:varepsilonPsi}), we expect the $\tcoso (\sigma)$ dependence to be quadratic:
		\begin{equation}
			T_\text{CO,SO}(\sigma_\text{[110]}) = T_\text{CO,SO}^{(0)} + \alpha_\text{eff.} \sigma_\text{[110]} + \beta_\text{eff.} \sigma_\text{[110]}^{2},
			\label{eq:tso110}
		\end{equation}
		where $\alpha_\text{eff.} = \partial T_\text{CO,SO} / \partial \sigma_\text{[100]}$ (i.e., exactly expression (\ref{eq:tso100})) and $\beta_\text{eff.}  = -4\beta / (G_\text{xy}^2 a)$. $\alpha_\text{eff.} $ and $\beta_\text{eff.} $ are effective parameters from $\alpha_1$, $\alpha_2$, and $\beta$ of the LFE model, with $G_\text{xy}$ denoting the in-plane shear modulus and $T_\text{CO,SO}^{(0)}$ the CO (or SO) transition temperature of the unstrained sample (see Appendix~D). The exact values of LFE expansion parameters $\alpha_1$, $\alpha_2$, and $\beta$ for either CO or SO are determined from the effective coefficients, and from experimental data, and rely on a precise quantification of the sample's elastic properties. \\
	\indent Using the elasticity data from \cite{NoharaPRB}, we apply the LFE model to our SO measurements by fitting (\ref{eq:tso100}) and (\ref{eq:tso110}) simultaneously, and yield (shown in Fig.~\ref{fig:model_fit}): 
	$\alpha_\text{eff.} ^\text{SO}= -\left(0.3 \pm 1.0\right)$~K/GPa, and  $\beta_\text{eff.} ^\text{SO} = -\left(21.3 \pm 4.0 \right)$~K/GPa$^2$ for the magnetic field aligned along the $c$ axis ([001]).\\ 
	\indent To test the validity of our model we wish to 
	use it to calculate the expected \tso\ suppression under hydrostatic pressure ($p$), and compare it to the values measured in Ref.~\cite{GuguchiaPRB}. 
	From the model it follows that $\tso (p)$ dependence is defined as:
	\begin{equation}
		\frac{\partial T_\text{SO}}{\partial p} = \frac{\partial T_\text{SO}}{\partial \sigma_\text{[100]}} -\frac{\alpha_1\nu_\text{out}}{Y_{\text{[100]}}a} + \frac{\alpha_2}{Y_{\text{[001]}}a}.
		\label{eq:tsohydro}
	\end{equation}
	The first term  (i.e., $\alpha_\text{eff.}$) characterizes the reaction to the in-plane symmetric  stress $\varepsilon_{A_\text{1g,1}}$, which we previously determined to be negligible.
	Once elastic constants and $\tso (\varepsilon)$ data are inserted, we can calculate that the expected hydrostatic suppression rate of $\frac{\partial T_\text{SO}}{\partial p} = -\left(3.9 \pm 2.1 \right)$~K/GPa (dashed green line in Fig.~\ref{fig:model_fit}). This value fits nicely to the comprehensive $\mu$SR dataset. Data for higher pressures were omitted for clarity. It should be kept in mind that the analysis is valid only until additional degrees of freedom, not accounted for in the model, start to contribute - e.g., interlayer coupling and suppression of the LTT phase with pressure.
	\\
	\indent Once the magnetic field is oriented along the [1\={1}0], it reduces the $T_\text{SO}(\sigma_\text{[110]})$ dependence drastically. From the LFE model-based symmetry point of view we can look at it in the following way: qualitatively, we expect the in-plane magnetic field  $H_\text{[1\={1}0]}$ to act on the $\Psi_{B_\text{2g}}$ magnetic order by breaking an additional symmetry. Therefore, the subsequent application of the in-plane stress is no longer symmetry-breaking, so the observed suppression of the $\tso$ is diminished. Overall, the effects of the magnetic field are two-fold: the increase in in-plane magnetization, which leads to non-vanishing Zeeman contribution to the free energy, and symmetry-breaking realized by the rotation of the in-plane magnetic moments \cite{HuckerPRB2008} through a spin-flop transition. The Zeeman contribution seems to be negligible since we do not observe a shift in \tso\ upon field rotation from $\text{[001]}$ to $\text{[1\={1}0]}$ at zero strain. To address the spin rotation, we utilize an atypic two-component order parameter represented just by the \blg\ and \bdg\ antisymmetric components:
	\begin{equation}
		\begin{pmatrix}
			\Psi_{B_\text{1g}} \\
			\Psi_{B_\text{2g}}
		\end{pmatrix} = 
		\begin{pmatrix}
			\Psi_{0} \cos(2\phi) \\
			\Psi_{0} \sin(2\phi)
		\end{pmatrix},
		\label{eq:doublecomponent}
	\end{equation}
	where $\Psi_{0} (H)$ is field-dependent order parameter magnitude, and angle $\phi$ describes a continuous rotation of the magnetic moments from the [100] and [010] directions to the [110] direction.
	To the lowest order in $\Psi$, this renormalizes the quadratic suppression coefficient $\beta_\text{eff.}$ upon applying [110] strain, while the behavior seems unchanged under symmetric strains.
	\\
	\indent It would be interesting to utilize our model and reproduce data in other systems. However, this is possible only if the complete strain data (for both [100] and [110] directions) are available. At the moment, only the present work has determined $\tso (\varepsilon_{[100]})$ and $\tso (\varepsilon_{[110]})$ dependencies. We can, nonetheless, note that the same model holds for the hydrostatic suppression of the CO  observed in Ref.~\cite{HuckerPRL}, and thus we can repeat the analysis to predict $\tco (p)$ using our $\tco(\varepsilon)$ data. As we have shown in Fig.~\ref{fig:phase_diagram}~b), the left-hand-side of expression (\ref{eq:tso100}) is again negligible, which (when combined with (\ref{eq:tso110})) leads to the following values of coefficients: $\alpha_\text{eff.} ^\text{CO} = -\left(1 \pm 5\right)$~K/GPa, and  $\beta_\text{eff.} ^\text{CO}= -\left(85 \pm 26 \right)$~K/GPa$^2$. The larger uncertainty here probably stems from a small number of points measured for $\tco(\varepsilon_{[100]})$ (the used sample was thicker than others, and thus maximum strain was limited by the maximum available stress our cell could apply). From these coefficients we calculate the expected behavior of $\tco(p)$ and show it as a dashed magenta line in Fig.~\ref{fig:model_fit}.  One should note, that the $\tco (p)$ data from Ref.~\cite{HuckerPRL} have large error bars for pressure values, which lead to larger uncertainty in determined $\tco$ values which is not shown in the figure. Nonetheless, the resulting curve follows the experimental data reasonably well.\\
	\indent The disappearance of CO in La$_{1.8-x}$Eu$_{0.2}$Sr$_{x}$CuO$_4$ has been viewed as entropy-driven~\cite{FinkPRB}, since LTT onsets so high that thermal energy destabilizes and melts the CO structure~\cite{PelcNC} before \tco\ approaches \tltt. In \LBCOc, \tltt\ does not increase with strain so if the same mechanism is at work it would mean that the LTT structure amplitude reduces with strain, which unpins the CO and thus suppresses \tco\ from the zero strain value of 54~K. This reduction of LTT amplitude is indeed seen in a recent work ~\cite{GuguchiaArXiv} on $x = 0.115$ doped sample, but it will require a separate study to check if it applies for the $1/8$-doped sample.\\ 
	\indent One cannot help but wonder how stress along [001] influences SO/CO. However, as such a study has various challenges, it is a topic for future work.
	\section{Conclusion}	
	\label{Con}
	\indent In summary, using $^{139}$La NMR relaxation rate $T_1 ^{-1}$ and $^{63}$Cu NQR spectra, we present the first study of phase diagrams of stripe spin order (SO), stripe charge order (CO) and LTT structure onset in \LBCOc, set by in-plane uniaxial strain ($\varepsilon$) in [100] and [110]. While the SO is more robust than at $x = 0.115$ doping, for $H\parallel [001]$ $\sigma_{[110]}$ stress dramatically suppresses \tso\ and no change is found for $\sigma_{[100]}$, which limits the applicability of theoretical models. Moreover, $H\parallel [1\overline{1}0]$ stabilizes the spin order. \\
	\indent CO shows the same response to strain as SO -- it is suppressed by $\varepsilon_{[110]}$ alone. The suppression decouples \tco\ and \tltt\ temperatures for $\varepsilon_{[110]} \ge 0.06\%$, and at maximum strain $\tltt -\tco$ even reaches 21~K. This separation reveals the role of symmetry in connecting two seemingly different doping phase diagrams - that of \LBCOx\ and   La$_{1.8-x}$Eu$_{0.2}$Sr$_{x}$CuO$_4$. Our results are understood using a symmetry-defined self-developed Landau free energy model that simultaneously shows a good agreement with existing data on hydrostatic $\tso (p)$ and $\tco (p)$ dependencies.
	
	\section{Acknowledgments}
	\indent The authors acknowledge the help of Christian Blum and Dina Bieberstein in preparing the samples, Sebastian Gass for SQUID measurements, and Markus H\"ucker and Ivan Kup\v ci\'c for valuable discussions. The work was supported by the Croatian Science Foundation (Grant No. IP-2018-01-2970), DFG (Grant No. DI2538/1-1), Alexander von Humboldt Foundation (Grant No. 3.4-1022249-HRV-IP),  and the project CeNIKS co-funded by the Croatian Government and the EU/ERDF - Competitiveness and Cohesion Operational Programme (Grant No. KK.01.1.1.02.0013). The work at BNL was supported by the US Department of Energy, Oﬃce of Basic Energy Sciences, contract No. DE-SC0012704. C.W.H. acknowledges support from DFG; TRR 288 - 422213477 (project A10).
	
	\section*{APPENDIX A: Technical details of the measurement setup}
	\begin{figure}[h]
		\includegraphics[width=0.4\textwidth]{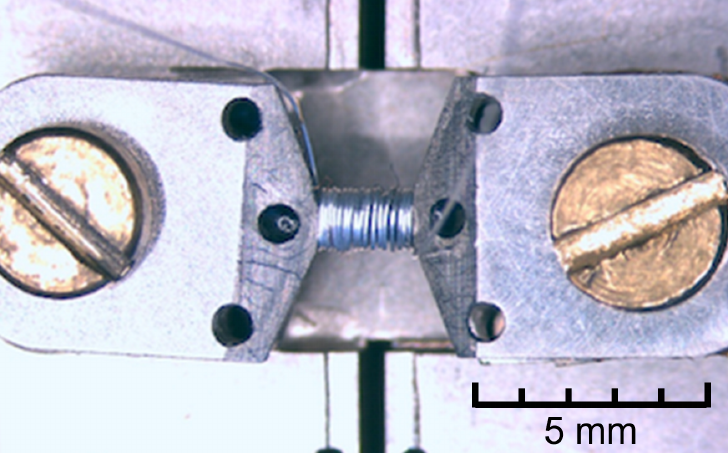}
		\caption{NMR coil with the sample in the uniaxial cell.}
		\label{fig:sample}
	\end{figure}
	In Fig.~\ref{fig:sample}, we show a part of our measurement setup with the sample and an NMR coil in a strain cell. The cell operation and strain analysis is described in~\cite{Hicks2014}. To gauge the uniaxial stress transferred to the sample, we used a simplified model:
	\begin{equation*}
		\sigma_\text{a} = \frac{Y_\text{a}\Delta L}{2\lambda+l_0},
	\end{equation*}
	where $Y_\text{a}$ is a Young modulus along a given axis, $\Delta L$ is a measured displacement change, and $l_0$ is the initial size of the sample along the strained dimension. The parameter $\lambda$, defining a length scale over which the stress is transferred to the sample, is given by:
	\begin{equation*}
		\lambda = \sqrt{\frac{Y_\text{a} t d}{2G}},
	\end{equation*}
	where $t$ and $d$ denote the thickness of the sample and epoxy, respectively, and $G$ is a shear strain modulus of the epoxy. We assume the epoxy to be an isotropic elastic material, and thus $G=Y_\text{epoxy}/(2+2\nu)$ where we take the Young modulus and Poisson ratio to be $Y_\text{epoxy} = 15$~GPa and $\nu= 0.3$ \cite{Hashimoto1980}. Unfortunately, the elastic constants for LBCO at 1/8 doping were not determined at cryogenic temperatures. However, data for similar compounds such as LSCO \cite{NoharaPRB} or LCO \cite{MiglioriPRL} corresponds to the transferred stress on the order of $\approx 1.5$~GPa at the highest applied voltages.
	At last, we have calculated the relative strain loss to the epoxy:
	\begin{equation*}
		\eta_\text{loss} = \frac{\Delta L - \Delta l_\text{sample}}{\Delta L} = \frac{2\lambda}{2\lambda + l_0},
	\end{equation*}
	which amounts to the loss $\eta_\text{loss} \approx 0.4-0.5$ for all our samples.
	
	\section*{APPENDIX B: NMR data acquisition information}
	\subsection*{Lanthanum spectra}
	
		\begin{figure}[h!]
		\includegraphics[width=0.5\textwidth]{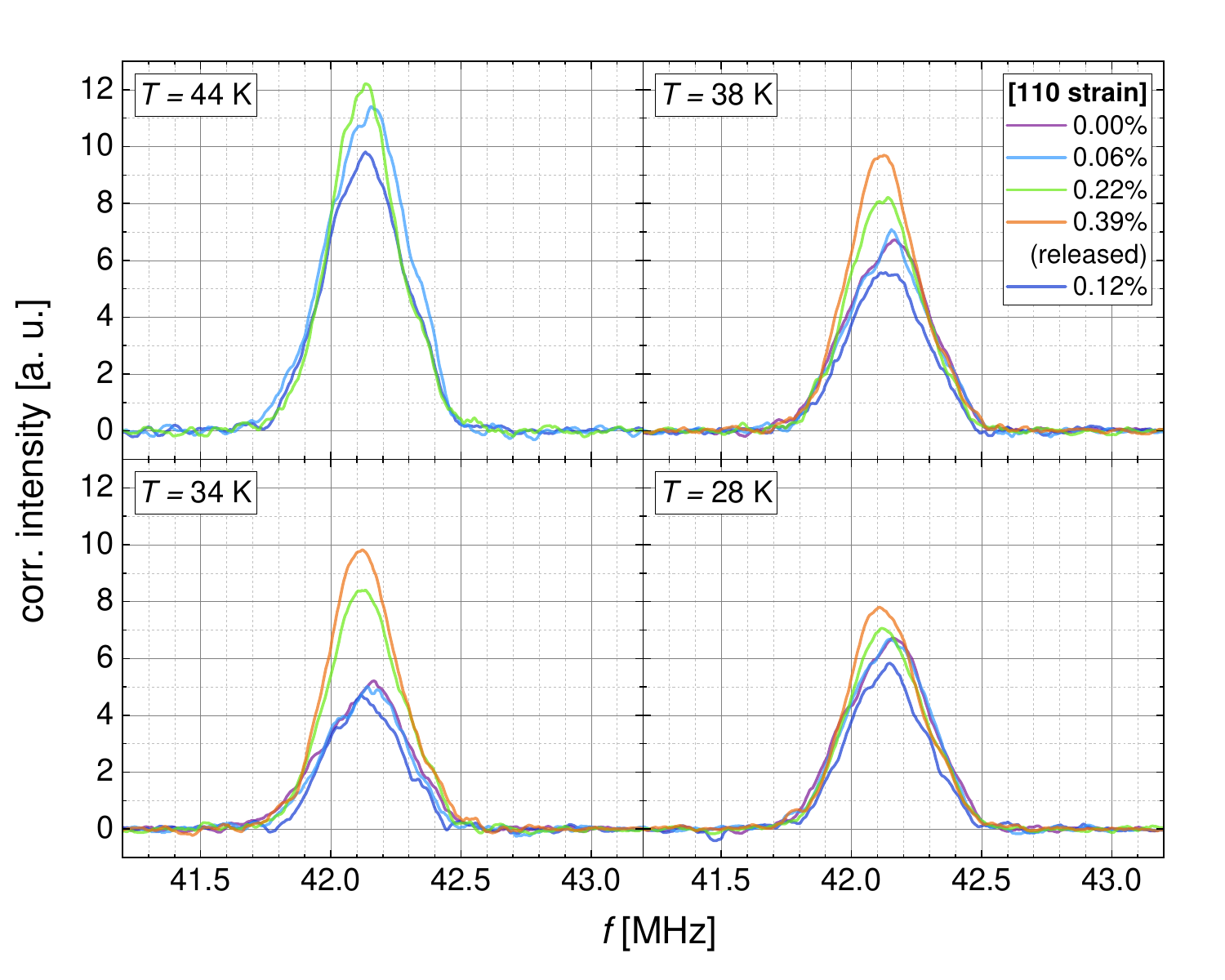}
		\caption{$^{139}$La NMR spectra (central transition) under different [110] uniaxial strains, at chosen temperatures above and below $T_\text{SO}$.
		}
		\label{fig:LaSpectra}
	\end{figure}
	
	In Fig. \ref{fig:LaSpectra}, we show temperature corrected $^{139}$La NMR spectra with uniaxial strain applied along [110] axis, for magnetic field along the [001] axis. We have observed no significant change in spectral width, and frequency with the applied [100] or [110] uniaxial strain. We attribute a noticeable decrease in the signal intensity across $T_\text{SO}$ to the enhanced longitudinal spin fluctuations near the spin-order transition. The spectra differ at the intermediate temperatures due to varying extent of the $T_\text{SO}$ suppression with the applied [110] uniaxial strain. At low temperatures ($T<28$~K), when spin fluctuations under different strains become comparable (Fig. 1 a)), the lineshapes coincide again. The effect is most noticeable at $T = 34$~K. Here, at low strains, the spectrum is measured precisely, or a bit below $T_\text{SO}$, and thus, the spectral intensity is significantly diminished. The change in the signal intensity is hardly noticeable at the highest strains, but becomes pronounced once again when the strain is released. When the strain is released, the original lineshape is recovered.

	\subsection*{Lanthanum relaxation curves}
\begin{figure*}[ht]
	\includegraphics[width=0.9\textwidth]{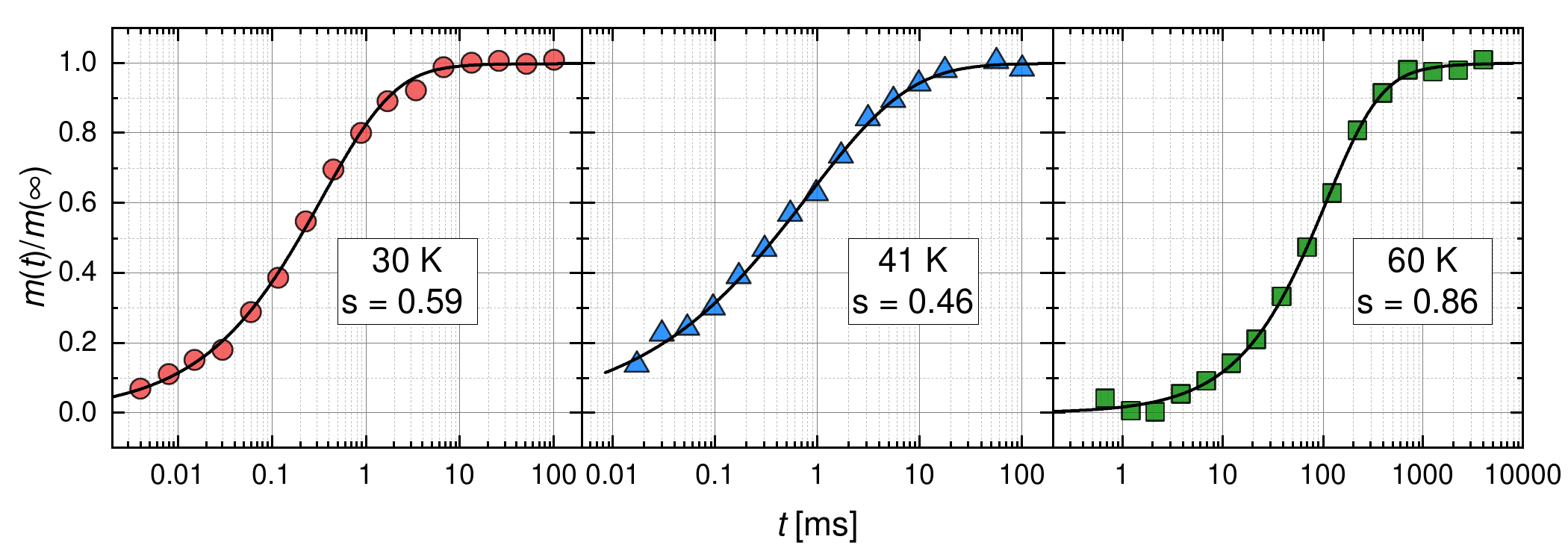}
	\caption{Stretching exponent $s$ fitted to our measurements. For clarity, we show only a subset of measured [110] strains with interpolated cubic splines as guides to the eye. Change in $s$ close to the transition temperature $T_\text{SO}$ is undoubtedly visible, and the strain dependence of the observed dip follows the same pattern as the $T_1$ data. The color-coded arrows mark \tso\ at respective strain.}
	\label{fig:LaRel}
\end{figure*}

	The fitting of spjn-lattice relaxation data of $^{139}$La central transition was done using the appropriate expression for the spin $ I = 7/2$ \cite{Narath1967,MacLaughlinPRB}: $f(t) = (1/84)e^{-(t/T_1)^{s}}  + (3/44) e^{-(6t/T_1)^{s}} + (75/364)e^{-(15t/T_1)^{s}} + (1225/1716)e^{-(28t/T_1)^{s}}$. The phenomenological stretching exponent $s$ gives insight into the distribution of the relaxation times $T_1$, as is explained in the main text.  In three panels of  Fig.~\ref{fig:LaRel}, we show the relaxation data and fits for zero strain, at temperatures of 30~K, 41~K and 60~K. The temperatures were selected to show the fit quality at three representative regimes of relaxation, that have different stretch exponent values, further discussed in the next paragraph.
	
\begin{figure}[h!]
	\includegraphics[width=0.5\textwidth]{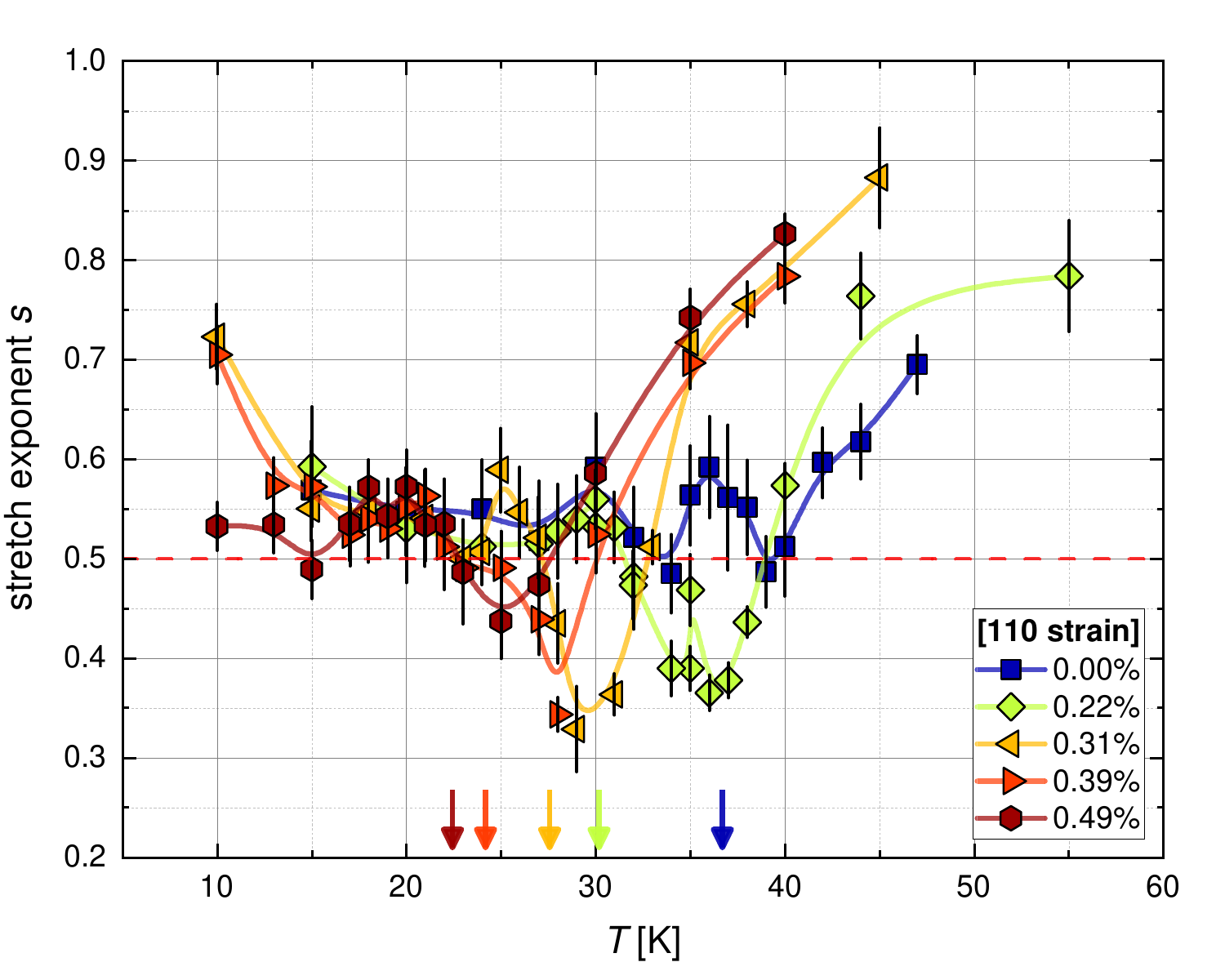}
	\caption{Stretching exponent $s$ fitted to our measurements. For clarity, we show only a subset of measured [110] strains with interpolated cubic splines as guides to the eye. Change in $s$ close to the transition temperature $T_\text{SO}$ is undoubtedly visible, and the strain dependence of the observed dip follows the same pattern as the $T_1$ data. The color-coded arrows mark \tso\ at respective strain.}
	\label{fig:LaStretch}
\end{figure}

	To accurately interpret the measured $T_1$ NMR relaxation data, we shall discuss the temperature and strain dependence of the fitted stretch exponent $s$ (Fig. \ref{fig:LaStretch}). When we approach the spin-order transition temperature $T_\text{SO}$ for a given strain, the $s$ dips abruptly. This behavior has already been observed in various cuprate systems, which exhibit a glassy type spin-order transition \cite{SingerPRB2020, Mitrovic2008}. We can see that the spatial distribution of $T_1$ times broadens significantly, but the stretch exponent stays predominantly larger than the threshold value of $s \approx 0.5$. It is, therefore, appropriate to analyze the fitted $T_1$ values as they always stay within $\sim 20\%$ of the distribution median. Conversely, it is justifiable to take a fixed value of $s$ to facilitate the interpretation of the fitted $T_1$ values \cite{CurroPRL2000}.
	
	\section*{Appendix C: Observing the LTO-LTT transition}
	\label{AppC}
		\begin{figure}[ht!]
		\includegraphics[width=0.5\textwidth]{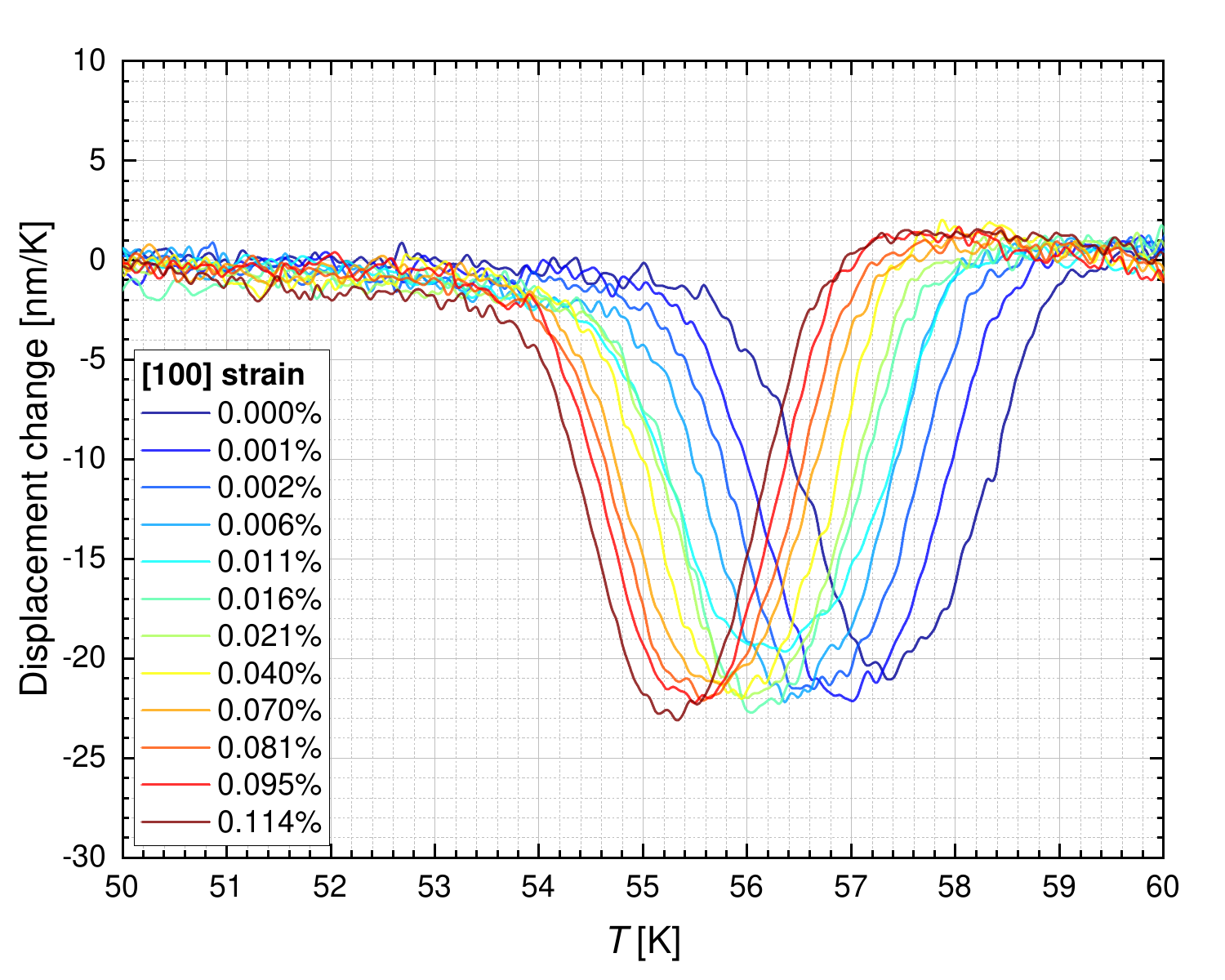}
		\caption{The anomaly in displacement change ($\Delta L / \Delta T$) of the strain cell measured in cooling ($r = 1$~K/min) for different applied stresses along [100] direction. The anomaly temperature coincides with the structural transition temperature $\tltt = 57.5$~K at zero applied stress. For increased strain values shown in the legend, the anomaly shifts to a lower temperature of 55.5~K.}
		\label{fig:displacement}
	\end{figure}
	Although the capacitive dilatometer of our strain cell has lower sensitivity than custom thermal-expansion measurement setups, it was sufficiently sensitive to detect a first-order LTO-LTT structural transition. We performed an exhaustive set of temperature sweeps at different uniaxial strains to characterize a change in the structural transition temperature $T_\text{LTT}$. We used two sweep rates, $r_1 = 1$~K/min and $r_2 = 0.5$~K/min, with each dataset measured for both cooling and warming, while the piezo stack voltage was held constant. Therefore, the observed displacement change should only come from the thermal expansion of the strain cell or the change in the sample's elastic properties. With the former being negligible in the measured temperature range, we can easily follow a structural transition as we increase the uniaxial stress on the sample.\\
	\indent When applying [110] uniaxial stress, the change in $T_\text{LTT}$ is absent or too small to be revealed by this method. In contrast, with the application of [100] stress sample displays a gradual, linear suppression of the $T_\text{LTT}$. Arguably, [100] stress promotes orthorhombicity and suppresses the transition to the LTT phase.\\
	\indent To confirm our dilatometry measurements, we look for the LTO-LTT structural transition in our $T_1$ NMR measurements. Using uniaxial stress along the [110] direction, we suppress the spin transition down to $\tso \approx 28$~K, revealing a discernible anomaly at $T_\text{LTT} \approx 56$~K which roughly coincides with the LTO-LTT transition. A similar feature was already observed in LESCO \cite{SimovicEPL} where the structural transition is separated from $\tso$ at zero strain. In addition to the slight increase in $T_1^{-1}$ relaxation rate, there is a discernible dip, LTT plateau in \ref{fig:LTT_T1}, in stretch exponent $s$ at $\tltt$, which implies a broader spatial distribution of the relaxation times $T_1$. This is consistent with the mixed-phase associated with the first-order structural transition.

	\section*{Appendix D: Calculation of the Landau free energy model}
	In the uniaxial strain experiment, it is advantageous to take the external stress applied on the sample as an independent variable. However, it is the induced strain that governs the suppression of the spin (charge) order transition temperature \tcoso\., so it is essential to handle the stress-strain conversion properly. In cuprates, and especially for LBCO and LSCO \cite{MiglioriPRL, NoharaPRB}, the elastic constants are given along the crystallographic axes of the high-temperature (HTT) phase. Suppose we wish to construct our free energy model in the LTT phase where the spin order sets in. In that case, we must transform the components of the stiffness matrix $\mathbf{C}$ using the familiar fourth-order tensor rotation formula:
	\begin{equation}
		C'_{ijkl} = c_i c_j c_k c_l C_{ijkl},
		\label{eq:4thTensor}
	\end{equation}
	where coefficients $c_i, c_j, c_k, c_l$ represent directional cosines along $i,j,k,l$ axes. In the transformation from the HTT to the LTT crystallographic axes, we can limit ourselves to the rotation about the $z$ axis ($\theta = \pm 45^\circ$). Equation \ref{eq:4thTensor} can then be condensed into a $6\times6$ rotation matrix:
	\begin{equation}
		\mathbf{R} = \begin{pmatrix}
			c^2 & s^2 & 0 & 0 & 0 & 2cs \\
			s^2 & c^2 & 0 & 0 & 0 & -2cs \\
			0 & 0 & 1 & 0 & 0 & 0 \\
			0 & 0 & 0 & c & s & 0 \\
			0 & 0 & 0 & -s & c & 0 \\
			-cs & cs & 0 & 0 & 0 & c^2-s^2 \\	
		\end{pmatrix},\quad \begin{matrix}
			c \equiv \cos{\theta} \\
			s \equiv \sin{\theta}
		\end{matrix},
	\end{equation}
	which acts on a stiffness tensor $\mathbf{C}^\text{(LTT)} = \mathbf{R}\mathbf{C}^\text{(HTT)}\mathbf{R}^\text{T}$. At last, to make the expressions more convenient to analyze and use, we replace the stiffness constant by utilizing the relation:
	\begin{align}
		&\left(\mathbf{C}^\text{(HTT,LTT)}\right)^{-1} = \mathbf{S}^\text{(HTT,LTT)} \nonumber\\
		&=
		\begin{pmatrix}
			\frac{1}{Y_\text{[100]}} & - \frac{\nu_\text{in}}{Y_\text{[100]}} & - \frac{\nu_\text{out}}{Y_\text{[100]}} & 0 & 0 & 0\\- \frac{\nu_\text{in}}{Y_\text{[100]}} & \frac{1}{Y_\text{[100]}} & - \frac{\nu_\text{out}}{Y_\text{[100]}} & 0 & 0 & 0\\- \frac{\nu_\text{out}}{Y_\text{[100]}} & - \frac{\nu_\text{out}}{Y_\text{[100]}} & \frac{1}{Y_\text{[001]}} & 0 & 0 & 0\\0 & 0 & 0 & \frac{1}{G_{zx}} & 0 & 0\\0 & 0 & 0 & 0 & \frac{1}{G_{zx}} & 0\\0 & 0 & 0 & 0 & 0 & \frac{1}{G_\text{xy}}
		\end{pmatrix}, 
	\end{align}
	where the elastic compliance matrix $\mathbf{S}$ is given in terms of Young and shear moduli ($Y_\text{[100]} = 233$~GPa, $Y_\text{[001]} = 176$~GPa, $G_{zx} \approx G_\text{xy} = 66.4$~GPa) and Poisson ratios ($v_\text{in} = 0.18$, $v_\text{out} = 0.27$). In this work, we use elastic stiffness constants given for the LTT phase when setting up the model but then express the results using the elastic parameters of the HTT lattice. The reason for this is twofold: the sample is oriented and glued into the strain cell with respect to the HTT axes, and we can readily use the elastic data from other sources to gauge the induced strain and expected \tcoso\ suppression. 
	
	To accentuate the role of the symmetry-breaking stress on the transition, we use an (anti) symmetrized strain components $\varepsilon_\algl = \frac{1}{2}\left(\varepsilon_{xx} + \varepsilon_{yy}\right)$, $\varepsilon_\algd = \varepsilon_{zz}$ and $\varepsilon_\blg = \frac{1}{2}\left(\varepsilon_{xx} - \varepsilon_{yy}\right), \varepsilon_\bdg = \varepsilon_\text{xy}$. From here, we construct a model taking into account five contributions to free energy:
 \begin{widetext}
	\begin{align}
		&F = F_\Psi + F_{\Psi\varepsilon} + F_{\Psi\Phi} + F_{\varepsilon} + F_{\sigma}, \nonumber\\
		&F_{\Psi} = \Psi_{B_\text{2g}}^{2} a \left(T - T_{SO}\right) + \Psi_{B_\text{2g}}^{4} b/2, \nonumber\\
		&F_{\Psi\varepsilon} =  \alpha_{1}\varepsilon_{A_\text{1g,1}} \Psi_{B_\text{2g}}^{2}  + \alpha_{2} \varepsilon_{A_\text{1g,2}} \Psi_{B_\text{2g}}^{2} +  \beta \varepsilon_{B_\text{1g}}^{2} \Psi_{B_\text{2g}}^{2}+ \gamma  \varepsilon_{B_\text{2g}} \Psi_{B_\text{2g}}, \nonumber\\
	 &F_{\Psi\Phi}^\text{(CO,SO)} = \delta \Phi_{B_\text{2g}}^2 \Psi_{B_\text{2g}}^2 ,\nonumber\\
	&F_{\varepsilon} = \varepsilon_{A_\text{1g,1}}^{2} \left(C_{11} + C_{12}\right) + C_{33} \varepsilon_{A_\text{1g,2}}^{2}/2 + \varepsilon_{B_\text{1g}}^{2} \left(C_{11} - C_{12}\right) +  2 C_{13} \varepsilon_{A_\text{1g,1}} \varepsilon_{A_\text{1g,2}}  + 2 C_{66} \varepsilon_{B_\text{2g}}^{2}, \nonumber\\
		 &F_{\sigma} = - \mathbf{\sigma} \cdot \mathbf{\varepsilon}
		\label{eq:LFEabstractmodel},
	\end{align}
	\end{widetext}
	where $\Psi_{B_\text{2g}}^{2}$ represents an emergent spin order which transforms as a $B_\text{2g}$ irreducible representation of a $D_\text{4h}$ point group, and $\Phi_{B_\text{2g}}$ a structural order parameter taken to be temperature independent for reasons listed in the article. All the contributions contain the lowest order terms in order parameters, with coupling constants expressed as $\alpha_1$, $\alpha_2$, $\beta$, $\gamma$ and $\delta$. $a$ and $b$ ($a, b >0$) are the standard Landau expansion parameters. The last, elastic energy contribution, sets the strains as a function of the applied uniaxial stress. At the minimum of the total free energy in the absence of the spin/structural order, $F_{\sigma}$ must be exactly equal to the quadratic form~in~strains~$F_{\varepsilon}$.\\
	\indent We can find the equilibrium strain as a solution to the set of minimization conditions $\frac{\partial F}{\partial \varepsilon_i} = 0$ given for all the symmetric and antisymmetric combinations of the strain. Evaluating the solution at $\sigma_\text{[100]} = 0$~GPa or $\sigma_\text{[110]} = 0$~GPa implies the emergence of spontaneous strains when the system enters an ordered phase:
	\begin{align}
		\varepsilon_{A_{1g;1}} &= \frac{\Psi_{B_{2g}}^{2} \left[(\nu_\text{in}-1) \alpha_{1} + 2 \nu_\text{out} \alpha_{2} \right]}{2 Y_\text{[100]}}, \nonumber\\
		\varepsilon_{A_{1g;2}} &= \frac{\Psi_{B_{2g}}^{2} \left(- Y_\text{[100]} \alpha_{2} + Y_\text{[001]} \nu_\text{out} \alpha_{1}\right)}{Y_\text{[100]} Y_\text{[001]}}, \nonumber\\
		\varepsilon_{B_{2g}} &= - \frac{\Psi_{B_{2g}} \gamma \left(\nu_\text{in} + 1\right)}{8 Y_\text{[100]}}, \nonumber\\  \varepsilon_{B_{1g}} &= \varepsilon_{E_g(1)} = \varepsilon_{E_g(2)} = 0.
	\end{align}
	Introduction of the equilibrium strain into the free energy model and minimization with respect to the order parameter $\Psi_{B_\text{2g}}$ results in a third order polynomial in $\Psi_{B_\text{2g}}$, with a single real solution. One may argue that the complex solutions to the order parameters are standard; however, we must disregard them as we have taken $\Psi_{B_\text{2g}}$ as the order magnitude, and we have allowed for a linear coupling in $\Psi_{B_\text{2g}}$. Therefore, such a solution would yield a non-physical complex free energy.
	
	The real solution for the stress $\sigma_{[100]}$ applied along [100] axis, implies that the \tcoso\ is suppressed in a linear fashion:
	\begin{equation}
		\frac{\partial \tcoso}{\partial \sigma_\text{[100]}} = \frac{(1-\nu_\text{in}) \alpha_{1}}{2 Y_\text{[100]} a} - \frac{\nu_\text{out} \alpha_{2}}{Y_\text{[100]} a} \equiv {f(\alpha_1, \alpha_2)}.
	\end{equation}
	Here, we observe that coupling constants $\beta$ and $\gamma$ are absent; thus, only the induced symmetric strains govern the suppression. We will encounter this expression multiple times, and therefore define it as a function $f(\alpha_1, \alpha_2)$. We purposefully consider $Y_\text{[100]}$ as a constant in $f(\alpha_1, \alpha_2)$ since the following expressions can always be expressed using exactly $Y_\text{[100]}$, irrespective of the direction of the applied stress. As noted in the article, we do not observe a measurable change in either \tso\ or \tco\ with this sample orientation, so that we can approximate $f(\alpha_1, \alpha_2) \approx 0$~K/GPa.
	When the stress $\sigma_{[110]}$ is applied to the sample, both \tso\ and \tco\ suppression rates are quadratic in $\sigma_\text{[110]}$. The linear term has the exact form as with the $\sigma_{[100]}$ stress, while the quadratic part depends on the sample's shear modulus $G_\text{xy}$:
	\begin{align}
		\Delta \tcoso(\sigma_\text{[110]}) &= f(\alpha_1, \alpha_2)\sigma_\text{[110]} - \frac{4 \beta}{G_\text{xy}^{2} a}\sigma_\text{[110]}^2 \nonumber\\
		&\approx  - \frac{4 \beta}{G_\text{xy}^{2} a} \sigma_\text{[110]}^2.
	\end{align}
	The \tcoso\ suppression under hydrostatic regime can also be expressed using $f(\alpha_1, \alpha_2)$, so we can reduce the dependence to:
	\begin{align}
		\frac{\partial \tcoso}{\partial p} &= f(\alpha_1, \alpha_2) - \frac{\nu_\text{out}\alpha_1}{Y_\text{[100]} a} + \frac{\alpha_2}{Y_\text{[001]} a}\nonumber\\
		 &\approx -\frac{\nu_\text{out}\alpha_1}{Y_\text{[100]} a} + \frac{\alpha_2}{Y_\text{[001]} a}.
		 \label{eq:Apphydro}
	\end{align}
	Note that the symmetric strain contribution $f(\alpha_1, \alpha_2)$ is present in both expressions for the \tcoso\ suppression rate. However, as discussed earlier, it seems to be negligible.
	
	Now, we turn our attention to the model extension, which describes the effect of the external magnetic field.
	The standard way of treating the in-plane external magnetic field is to include a Zeeman contribution $F_\text{Zeeman} = \mu \mathbf{H} \cdot \mathbf{m}(\Psi_{B_{2g}})$, where $\mathbf{m}(\Psi_{B_{2g}})$ represents a magnetic moment associated with the order parameter $\Psi_{B_\text{2g}}$. Unfortunately, it is immediately evident that such a contribution would lead to a change in \tso\ at all strains. 
	In our model, we propose a two-component order parameter by introducing in-plane order parameters which are defined by different symmetry properties: $\Psi_{B_\text{1g}}$ transforms as $B_\text{1g}$, and $\Psi_{B_\text{2g}}$ transforms as ${B_\text{2g}}$ representation of the $D_\text{4h}$ point group. We proceed to write down the Landau model in the absence of strain up to the fourth-order invariants:
	\begin{align*}
		F_\Psi &=  a \left(T - T_{SO}\right) (\Psi_{B_{1g}}^{2}+\Psi_{B_{2g}}^{2}) + \frac{b (\Psi_{B_{1g}}^{4} + \Psi_{B_{2g}}^{4})}{2} +\nonumber\\
		&+ c\Psi_{B_{1g}}^{2}\Psi_{B_{2g}}^{2}.
	\end{align*}
	
	Here, we realize that the assumption $c \approx b$ allows for a convenient reparametrization of the order parameters:
	\begin{equation}
		\begin{pmatrix}
			\Psi_{B_\text{1g}} \\
			\Psi_{B_\text{2g}}
		\end{pmatrix} = 
		\begin{pmatrix}
			\Psi_{0} \cos(\varphi) \\
			\Psi_{0} \sin(\varphi)
		\end{pmatrix},
		\label{eq:doublecomponentabstract}
	\end{equation}
	where $\Psi_0$ represents a total order magnitude and $\varphi$ an angle that defines the mixing of the two components. The minimization of the proposed Landau model with respect to $\Psi_0$ determines that the spin order $\Psi_0 = \sqrt{a(T-\tso)/b}$ sets in strictly at \tso\ irrespective of the component mixing angle $\varphi$.
	The crucial difference from the single component model is that we must include all the strain-coupling to the lowest order of $\Psi_{B_{1g}}$ and $\Psi_{B_{1g}}$:
	\begin{align}
		F_{\Psi_{B_{1g}}} &= \alpha_{11} \varepsilon_{A_{1g;1}} \Psi_{B_{1g}}^{2}  + \alpha_{21} \varepsilon_{A_{1g;2}} \Psi_{B_{1g}}^{2} +\nonumber\\
		&+ \beta_{21} \varepsilon_{B_{2g}}^{2} \Psi_{B_{1g}}^{2} + \beta_{11} \varepsilon_{B_{1g}} \Psi_{B_{1g}}, \\
		F_{\Psi_{B_{2g}}} &= \alpha_{12} \varepsilon_{A_{1g;1}} \Psi_{B_{2g}}^{2}  +  \alpha_{22} \varepsilon_{A_{1g;2}} \Psi_{B_{2g}}^{2} +\nonumber\\
		&+  \beta_{12} \varepsilon_{B_{1g}}^{2} \Psi_{B_{2g}}^{2} + \beta_{22} \varepsilon_{B_{2g}} \Psi_{B_{2g}},
	\end{align}
	where coefficients $\alpha_{ij}$ define coupling strength to the symmetric, and $\beta_{ij}$ to the asymmetric strain (we take the first index $i$ to refer to the strain component, e.g., $i=1 \rightarrow \varepsilon_{A_{1g,1}}$, and second index $j$ to refer to the symmetry of the order parameter). The spin-structure coupling and the elastic energy contribution are left unchanged.
	
	With the introduction of the order parameter reparametrization and the minimization of the free energy with respect to $\Psi_0$, in case of the $\sigma_{[110]}$ strain, we get:
	\begin{align}
	&	\Delta \tso(\sigma_{[110]}, \varphi) =\left[f(\alpha_{11}, \alpha_{21})\sin^2{\varphi} +\right. \nonumber\\
	&\left. + f(\alpha_{12}, \alpha_{22})\cos^2{\varphi}\right]\sigma_{110} - \frac{\beta_{12} \cos^{2}{\left(\varphi \right)}}{G_\text{xy}^{2} a}\sigma_{110}^{2} .
	\end{align}
	We have already demonstrated that the suppression rate $f(\alpha_{12}, \alpha_{22})$, related to the $\Psi_{B_{2g}}$ spin order, vanishes, but one should not assume the same for $f(\alpha_{11}, \alpha_{21})$ rate. Nevertheless, by fitting the quadratic function to our measurements, we can show that the quadratic suppression constant $\beta_\text{eff.}$ is indeed reduced by some factor $\cos^2{\varphi}$.
	Finally, more experimental data is needed to get the exact dependence of the mixing angle $\varphi$ on the orientation of the applied in-plane magnetic field. However, by looking at the crystal symmetry, we must assume that the model is symmetric to rotation by $\phi = 90^\circ$ when the spin stripe direction coincides again with CuO bonds. In order to correlate the model to the structure, in the main article we use the reparametrization with $\varphi = 2\phi$.
	
	\bibliographystyle{apsrev}
	\bibstyle{prb}

\end{document}